\documentclass[sigconf]{acmart}
\usepackage{amsmath}
\usepackage{color}
\usepackage{array}
\usepackage{breqn}
\usepackage{{inputenc}}
\usepackage{balance}
\usepackage{multirow,tabularx}
\usepackage{booktabs,longtable}
\usepackage{hhline}
\usepackage[flushleft]{threeparttable}
\usepackage{algorithm}
\usepackage{algorithmic}
\usepackage{epsfig}
\usepackage{graphicx}
\usepackage{epstopdf}
\usepackage{thmtools}
\usepackage{enumitem}
\usepackage{verbatim}
\usepackage{amsfonts}
\usepackage{hyperref}
\hypersetup{colorlinks=false,linkcolor=blue,urlcolor=blue,citecolor=red}
\usepackage{xcolor}
\usepackage{subcaption}
\usepackage{wrapfig}
\usepackage{makecell}
\usepackage{bbm}

\usepackage{tcolorbox}
\usepackage{xcolor}
\usepackage{svg}
\usepackage{bm}
\usepackage{afterpage}
\usepackage{mdframed}
\tcbuselibrary{skins, breakable}

\newsavebox{\SummaryGeneration}
\newsavebox{\TasteGeneration}
\newsavebox{\ReactToItems}
\newsavebox{\SatReflect}
\newsavebox{\NextDecision}
\newsavebox{\Examples}
\newsavebox{\InterviewCase}

\newcommand{\Set}[1]{\mathcal{#1}}

\newcommand{\ie}{\emph{i.e., }}
\newcommand{\eg}{\emph{e.g., }}

\newcommand{\cf}{\emph{cf. }}

\newcommand{\keys}[1]{{\color{black}{#1}}}

\newcommand{\slh}[1]{{\color{black}{#1}}}

\newcommand{\cmmnt}[1]{}

\definecolor{format}{RGB}{148, 99, 24}
\definecolor{user}{RGB}{169, 16, 3}
\definecolor{item}{RGB}{15, 86, 157}

\AtBeginDocument{%
  \providecommand\BibTeX{{%
    \normalfont B\kern-0.5em{\scshape i\kern-0.25em b}\kern-0.8em\TeX}}}

\copyrightyear{2024}
\acmYear{2024}
\setcopyright{rightsretained}
\acmConference[SIGIR '24]{Proceedings of the 47th International ACM SIGIR
Conference on Research and Development in Information Retrieval}{July 14--18,
2024}{Washington, DC, USA}
\acmBooktitle{Proceedings of the 47th International ACM SIGIR Conference on
Research and Development in Information Retrieval (SIGIR '24), July 14--18,2024, Washington, DC, USA}
\acmDOI{10.1145/3626772.3657844}
\acmISBN{979-8-4007-0431-4/24/07}

\makeatletter
\gdef\@copyrightpermission{
  \begin{minipage}{0.3\columnwidth}
   {\includegraphics[width=0.90\textwidth]{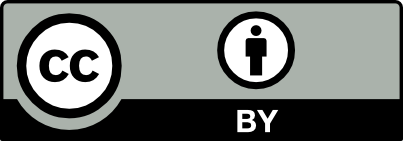}}
  \end{minipage}\hfill
  \begin{minipage}{0.7\columnwidth}
   \href{https://creativecommons.org/licenses/by/4.0/}{This work is licensed under a Creative Commons Attribution International 4.0 License.}
  \end{minipage}
  \vspace{5pt}
}
\makeatother

 \hypersetup{hidelinks,
 	colorlinks=true,
 	allcolors=black,
 	pdfstartview=Fit,
 	breaklinks=true}

\begin{document}

\definecolor{background_u}{HTML}{FEF9F5}
\definecolor{frame_u}{HTML}{98450E}
\definecolor{background_i}{HTML}{F9FBFD}
\definecolor{frame_i}{HTML}{2E75B5}
\definecolor{background_e}{HTML}{FAFAFA}
\definecolor{frame_e}{HTML}{0D0D0D}

% \title{Learning disentangled representations via latent space data augmentation for Collaborative Filtering}
\title{On Generative Agents in Recommendation}

%\email{webmaster@marysville-ohio.com}
%\affiliation{%
%  \institution{Institute for Clarity in Documentation}
%  \streetaddress{P.O. Box 1212}
%  \country{USA}

\author{An Zhang$^{*}$}\thanks{$^{*}$ equal contribution}
\affiliation{%
  \institution{National University of Singapore} \country{Singapore}
    % \city{Singapore}
  \country{}}
\email{anzhang@u.nus.edu}

\author{Yuxin Chen$^{*}$}
\affiliation{%
  \institution{National University of Singapore}
  \country{Singapore}
    % \city{Singapore}
  \country{}}
\email{e1143404@u.nus.edu}

\author{Leheng Sheng$^{*}$}
\affiliation{%
  \institution{Tsinghua University}
  \city{Beijing}
  \country{China}
}
\email{chenglh22@mails.tsinghua.edu.cn}

\author{Xiang Wang$^{\dag}$}\thanks{$^{\dag}$Xiang Wang is the corresponding author, also affiliated with the Institute of Dataspace, Hefei Comprehensive National Science Center.}
\affiliation{%
  \institution{University of Science and Technology of China}
    \city{Hefei}
  \country{China}}
\email{xiangwang1223@gmail.com}

\author{Tat-Seng Chua}
\affiliation{%
  \institution{National University of Singapore}
  % \country{Singapore}
    % \city{Singapore}
  \country{Singapore}}
\email{dcscts@nus.edu.s}

% \author{\textbf{An Zhang}}
% \affiliation{%
%   \institution{National University of Singapore}
%   \institution{Sea-NExT Joint Lab}
%   \country{Singapore}}
% \email{anzhang@u.nus.edu}

% \author{\textbf{Jingnan Zheng}}
% \affiliation{%
%   \institution{National University of Singapore}
%   \country{Singapore}}
% \email{e0718957@u.nus.edu}

% \author{\textbf{Xiang Wang$^\S$}}\thanks{$^\S$Xiang Wang is corresponding author, also with Institute of Artificial Intelligence, Hefei Comprehensive National Science Center.}
% \affiliation{%
%   \institution{University of Science and Technology of China}
% %   \institution{Institute of Artificial Intelligence, Hefei Comprehensive National Science Center}
%   \country{China}
%   }
% %   \institution{Sea-NExT Joint Lab}}
% %   \institution{Sea-NExT Joint Lab}
% %   \streetaddress{streat address}
% %   \city{city}
% %   \country{Singapore}}
% \email{xiangwang1223@gmail.com}

% \author{\textbf{Tat-Seng Chua}}
% \affiliation{%
%   \institution{National University of Singapore}
%   \institution{Sea-NExT Joint Lab}
%   \country{Singapore}}
% \email{dcscts@nus.edu.sg}

\renewcommand{\shortauthors}{An Zhang, Yuxin Chen, Leheng Sheng, Xiang Wang, and Tat-Seng Chua}

\begin{abstract}
Recommender systems are the cornerstone of today's information dissemination, yet a disconnect between offline metrics and online performance greatly hinders their development.
Addressing this challenge, we envision a recommendation simulator, capitalizing on recent breakthroughs in human-level intelligence exhibited by Large Language Models (LLMs).
We propose \textbf{Agent4Rec}, a user simulator in recommendation, leveraging LLM-empowered generative agents equipped with user profile, memory, and actions modules specifically tailored for the recommender system.
In particular, these agents' profile modules are initialized using real-world datasets (\eg MovieLens, Steam, Amazon-Book), capturing users' unique tastes and social traits; memory modules log both factual and emotional memories and are integrated with an emotion-driven reflection mechanism; action modules support a wide variety of behaviors, spanning both taste-driven and emotion-driven actions.
Each agent interacts with personalized recommender models in a page-by-page manner, relying on a pre-implemented collaborative filtering-based recommendation algorithm.
We delve into both the capabilities and limitations of Agent4Rec, aiming to explore an essential research question: ``To what extent can LLM-empowered generative agents faithfully simulate the behavior of real, autonomous humans in recommender systems?''
Extensive and multi-faceted evaluations of Agent4Rec highlight both the alignment and deviation between agents and user-personalized preferences.
Beyond mere performance comparison, we explore insightful experiments, such as emulating the filter bubble effect and discovering the underlying causal relationships in recommendation tasks. 
Our codes are available at \url{https://github.com/LehengTHU/Agent4Rec}.
% Our codes are available at \url{https://anonymous.4open.science/r/Agent4Rec-2BFB/}.
\end{abstract}





\begin{teaserfigure}
      \centering
      \includegraphics[width=0.8\linewidth]{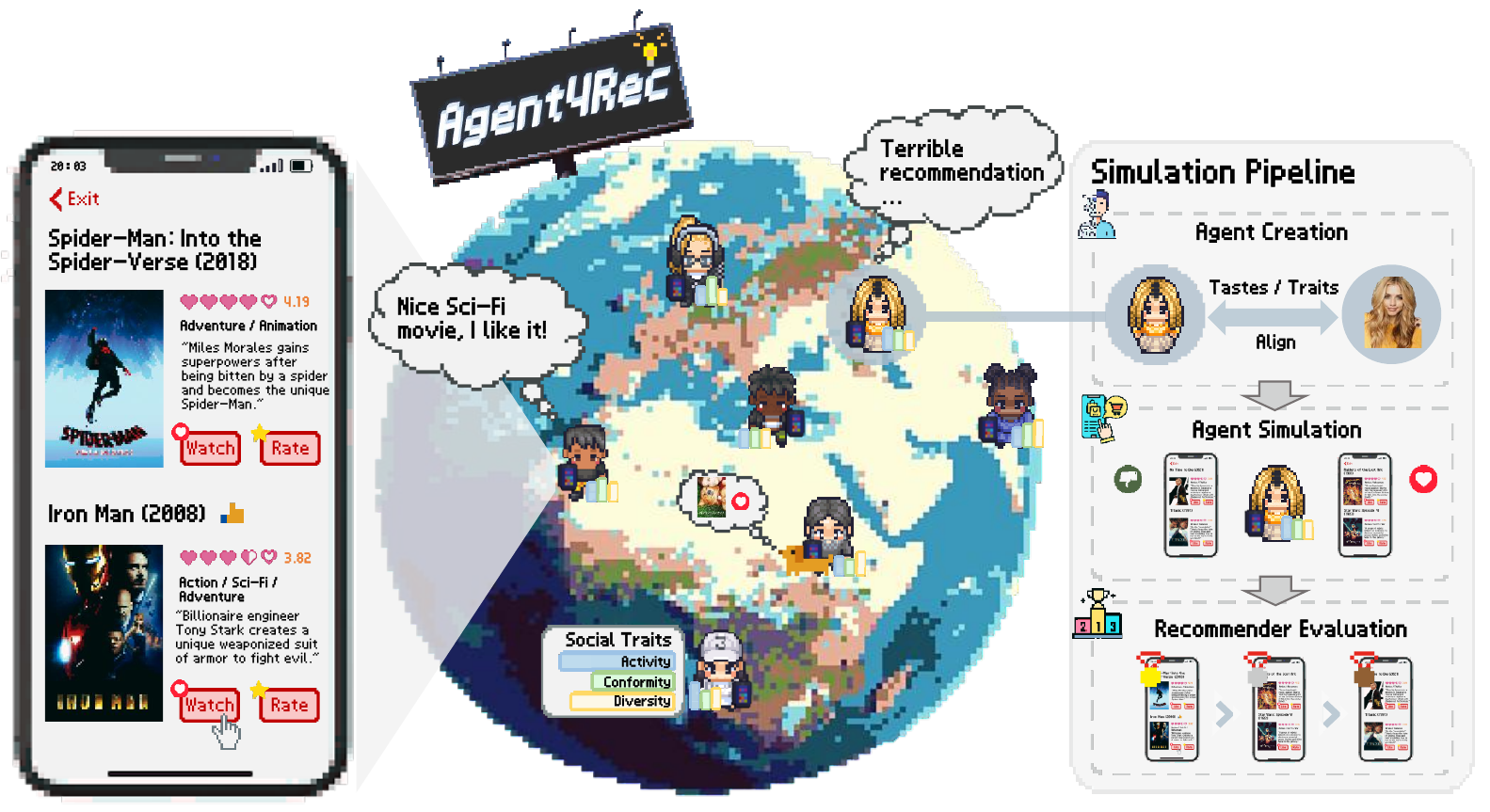}
      \vspace{-10pt}
      \caption{Illustration of Agent4Rec, a user simulator with 1,000 LLM-empowered generative agents in the movie recommendation scenario.
      These agents are initialized from the MovieLens-1M \cite{MovieLens} dataset, embodying varied social traits and preferences. 
      Each agent interacts with personalized movie recommendations in a page-by-page manner and undertakes various actions such as watching, rating, evaluating, exiting, and interviewing.
      % With Agent4Rec, we delve into a pioneering exploration to address multi-faceted challenges: How closely do agents align with genuine human preferences? Which recommendation algorithms excel in performance and under what metrics? What are the unresolved problems in recommender systems, and how might they be mitigated using simulation-driven insights?
      % With the specialized simulator tailored for recommendations, Agent4Rec, we would like to explore an essential question: To what extent can LLM-empowered generative agents truly simulate the behavior of genuine, independent humans in recommender systems? 
      % Moreover, can such simulation provide some insights into unresolved questions? 
      With Agent4Rec, we would like to explore the potential of LLM-empowered generative agents in simulating the behavior of genuine, independent humans in recommendation environments. 
      % Additionally, we hope that this simulation can offer insights into some unresolved questions.
      }
      \label{fig:sandbox}
      % \vspace{10pt}
\end{teaserfigure}

\maketitle

\section{Introduction}
Recommender systems play a pivotal role in contemporary information dissemination, actively shaping individual preferences and cognitive processes \cite{bc_loss, llara}. 
Despite their great success and widespread adoption, the conventional supervised recommendation approach falls short, as evidenced by the significant gap between offline metrics and online performance \cite{RecSys_survey, AdvInfoNCE}.
This disconnect hinders the integration of academic research into real-world recommendation deployments, acting as a bottleneck for the field's future advancements \cite{Netflix, ABtest}.
Imagine a paradigm where a configurable simulation platform for recommender systems exists --- one that faithfully captures user intent and encodes human cognitive mechanisms.
Such a simulator undoubtedly has the potential to revolutionize traditional research paradigms in recommendations, offering an innovative pathway for data collection, recommender evaluation, and algorithmic development \cite{challenges_rl, Virtual-Taobao, RecSim}. 

Recent strides in Large Language Models (LLMs), with their impressive capability \cite{deductive_reasoning, Survey_Fudan, InstructGPT} and profound comprehension of time and space \cite{LLM_space, Survey_Renmin, CoT}, underscore the promise of the recommendation simulator paradigm.
Specifically, LLMs serve as the foundational architecture in the development of generative agents \cite{Generative_Agent, ChatDev, S3}.
These agents are then integrated into a recommendation environment, taking on the virtual users in simulators.
% To ensure a faithful emulation of user behavior, these agents are enhanced through the incorporation of designed modules, such as memory and action components, replicating essential cognitive mechanisms \cite{ChatDev, S3}.
Yet, in this promising research direction, developing a reliable simulator that faithfully mirrors personalized user preferences is non-trivial \cite{toolrec, RecAgent}. 
Consequently, harnessing the potential of LLM-empowered generative agents by designing modules tailored for recommendation to emulate human behavior has become a key research focus.
% How to fulfill the potential of LLM-empowered generative agents by designing specifically tailored recommendation modules is a pressing and critical challenge.
% \begin{itemize}[leftmargin=*]
%     \item How to build? The creation of \keys{a reliable simulation that faithfully mirrors personalized user behavior} is non-trivial \cite{RecAgent}.
%     It requires generative agents to understand diverse user social traits and individual preferences.
%     \item How to evaluate? Determining the extent to which a simulator accurately reflects user preference and evaluates the recommendation algorithms poses another formidable challenge \cite{AgentBench}.
%     \item How to fully utilize? The vast potential of recommendation simulators, especially in addressing the unresolved problems in the recommendation domain, remains to be thoroughly investigated. 
% \end{itemize}
% However, a substantial challenge that remains unaddressed is the creation of a believable simulation of personalized user actions, accounting for diverse user social traits and individual preferences. 
% Furthermore, the systematic evaluation of these simulation platforms and their potential to shed light on unresolved issues in recommendation tasks has yet to be explored.

\begin{figure*}[t]
    % \vspace{10pt}
    \includegraphics[width=0.85\linewidth]{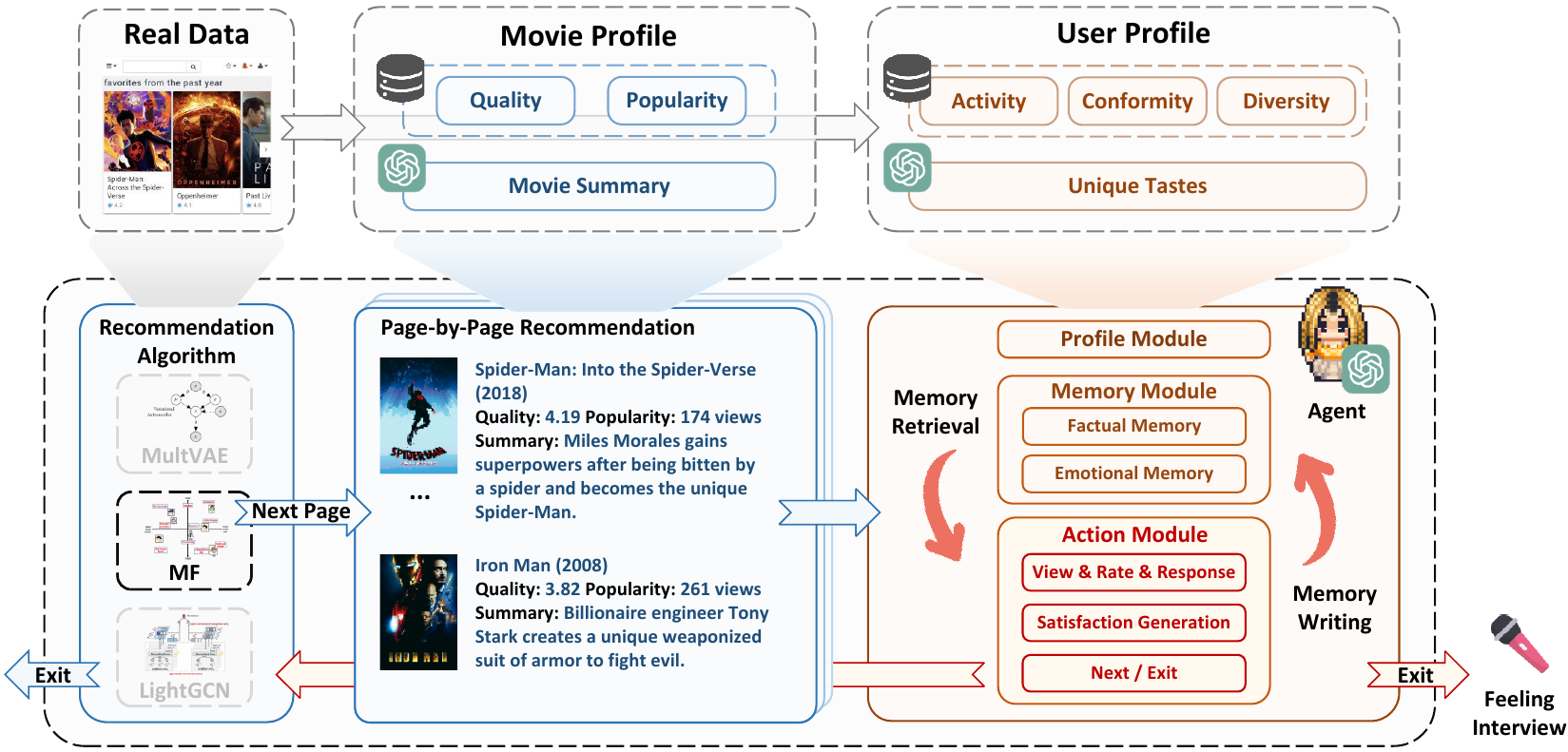}
    \vspace{-5pt}
    \caption{The overall framework of Agent4Rec.
    Our simulator consists of two core facets: LLM-empowered Generative Agents (Red Section) and Recommendation Environment (Blue Section).
    Both user and item profiles are initialized using real-world datasets (\eg MovieLens-1M, Steam, Amazon-Book).
    The recommendation algorithm, an adaptable component of the system, generates item recommendations presented to the agents in a page-by-page manner.
    Agents, enhanced with specialized memory and action modules tailored for recommendation scenarios, simulate a wide range of behaviors including viewing, rating items, providing feedback, navigating to the next page, exiting the system, and participating in post-exit interviews.
    }
    \vspace{-5pt}
    \label{fig:framework}
\end{figure*}

To bridge this gap, we introduce \textbf{Agent4Rec} --- a general user simulator in recommendation scenarios, which consists of two core facets: LLM-empowered generative agents and recommendation environment (\cf Figure \ref{fig:framework}). 
From the user's perspective, we simulate 1,000 LLM-empowered generative agents per recommendation scenario, where each agent is initialized based on real-world datasets and composed of three essential modules: the user profile, memory, and action modules.
The profile module functions as a repository for personalized social traits and historical preferences \cite{proflile_survey}, facilitating the alignment of user portraits with genuine human characteristics.
The memory module records past viewing behaviors, system interactions, and emotional memories (\ie user feelings and fatigue levels) in natural language, enabling information retrieval, preference accumulation, and emotion-driven reflection in a coherent manner.
The action module empowers these agents to interact directly with the recommendation environment, including taste-driven actions (\ie viewing or ignoring recommended movies, rating, generating post-viewing feelings), and emotion-driven actions (\ie exiting the system, evaluating recommendation lists, and expressing human-understandable comments).
From the perspective of the recommender system simulation, items are recommended by a predetermined recommendation algorithm, sequenced in a page-by-page format to emulate real-world recommendation platforms.
% the \keys{recommendation algorithm} is an independent entity while being designed to work in conjunction with the simulator. 
On the one hand, our simulator predominantly integrates collaborative filtering-based recommendation strategies, encompassing random, most popular, Matrix Factorization (MF) \cite{MF}, LightGCN \cite{LightGCN}, and MultVAE \cite{MultVAE}.  
On the other hand, we architect the simulator with extensibility as a core principle. 
By providing open interfaces, we empower researchers and practitioners to effortlessly deploy any recommendation algorithm of their choice. 
% This ensures that, in addition to the integrated strategies, our simulator serves as a versatile sandbox for comprehensive evaluation and the collection of valuable user feedback.

To systematically evaluate the effectiveness and limitations of our proposed Agent4Rec, we conduct comprehensive experiments from both the user's and recommender system's perspectives.
From the user's standpoint, our primary focus lies in assessing \keys{the degree of agent alignment}. 
Specifically, we evaluate to what extent the agent can ensure the coherence of the \keys{true user's social traits, personality, and preferences} using a variety of metrics and \keys{statistical tests}.
On the recommender system simulation front, we evaluate various recommenders configured with different algorithms.
Our evaluation metrics span multiple dimensions, including the average number of items recommended and watched by users, average user ratings, user engagement time, and overall user satisfaction.
In parallel, the agent feedback collected from the simulator serves as augmented data, enabling iterative training and refinement of the recommendation strategies. 
Subsequently, we assess the feedback-driven recommender enhancement using standardized offline metrics. 
This dual approach, combining simulation feedback with traditional offline evaluation, ensures a comprehensive assessment of recommendation algorithms.

To explore the potential of the simulator in investigating unresolved challenges in recommendation tasks, we undertake two experiments.
In the first experiment, we emulate the filter bubble effect within the simulator --- a scenario where users are consistently exposed to similar or reinforcing content, resulting in a reduction in item attribute diversity \cite{rl_filter_bubble, CIRS}. 
This investigation aims to understand the extent to which feedback loops can amplify such centralized recommendation phenomena.
Additionally, we utilize the simulator as a data collection tool to pioneer a data-oriented causal discovery \cite{CausPref, ReScore}. 
This approach yields a robust causal graph, enabling us to unveil the intricate latent causal relationships that infer the data generation progress in recommender systems \slh{\cite{causal_survey}}.

Our main contributions are summarized as follows:
% \vspace{-3pt}
\begin{itemize}[leftmargin=*]
    \item We develop Agent4Rec, a general recommendation simulator utilizing LLM-empowered agents to emulate and deduce user-personalized preferences and behavior patterns. 
    These agents, with their carefully designed modules tailored to recommendation, enable the emulation of human cognitive mechanisms.
    \item We delve into both the capabilities and limitations of Agent4Rec by conducting extensive evaluations for generative agent-based simulation in recommender systems.
    We employ statistical metrics and tests for user alignment evaluation and propose a dual parallel evaluation considering both offline performance and simulation feedback. 
    \item Using Agent4Rec as a data collection tool, we replicate a pervasive issue --- the filter bubble effect—and unveil the underlying causal relationships embedded within recommender system scenarios. 
    These observations showcase the potential of Agent4Rec to inspire new research directions.
\end{itemize}

We believe that Agent4Rec stands at the intersection of cutting-edge technology and the challenges of recommender systems, offering an experimental platform for insights that will inspire more work in this research direction.

\section{Agent4Rec}
Agent4Rec, as a user simulation in recommendation scenarios, is expected to accurately mirror user behaviors, effectively forecast long-term user preferences, and systematically evaluate recommendation algorithms by leveraging the human-like capabilities of LLM-empowered generative agents.
To achieve this goal, two core facets are considered: (1) designing agent architectures that faithfully mimic user personalized preferences and human cognitive reasoning, and (2) constructing a recommendation environment that ensures its reliability, extensibility, and adaptability.
Figure \ref{fig:framework} demonstrated the overview of Agent4Rec’s framework, which is developed by modifying \href{https://www.langchain.com/}{LangChain}, with all agents being powered by the gpt-3.5-turbo version of ChatGPT.

\smallskip\noindent\textbf{Task Formulation.} \label{sec:task-formulation}
% The \slh{recommender} system simulator aims to \note{construct a platform that accurately mirrors user behaviors, forecasts long-term user preferences and optimizes recommendation benefits.}
Given a user $u \in \Set{U}$ and an item $i \in \Set{I}$, let $y_{ui} = 1$ denote that user $u$ has interacted with item $i$, and subsequently rated it with $r_{ui} \in \{1,2,3,4,5\}$. 
Conversely, $y_{ui} = 0$ indicates that the user has not adopted the item.
The quality of each item $i$ can be represented by $R_i \doteq \frac{1}{\sum_{u \in \Set{U}}y_{ui}}\sum_{u \in \Set{U}} y_{ui} \cdot r_{ui}$, while its popularity is denoted by $P_i$. Additionally, the genre set of the item is given by $G_i \subset \Set{G}$.
The simulator's overarching goal is to faithfully distill the human genuine preferences such as $\hat{y}_{ui}$ and $\hat{r}_{ui}$ of user $u$ for an unseen recommended item $i$.

\subsection{Agent Architecture}
Generative agents in Agent4Rec, utilizing LLM as its foundational architecture, refine their capabilities with three specialized modules tailored for recommendation scenarios: a profile module, a memory module, and an action module.  
Specifically, to emulate personalized genuine human behaviors, each agent integrates a user profile module to reflect individualized social traits and preferences.
Furthermore, drawing inspiration from human cognitive processes, agents are equipped with memory and action modules, enabling them to store, retrieve, and apply past interactions and emotions to generate behaviors in a coherent manner.  

\subsubsection{\textbf{Profile Module.}}\label{sec:method_profile}
In the domain of personalized recommendation simulation, the user profile module stands as a cornerstone, playing a crucial role in the efficacy of agents' alignment with genuine human behaviors.
To lay a reliable foundation for the generative agent's subsequent simulations and evaluations, the benchmark dataset (\eg MovieLens-1M \cite{MovieLens}, Steam \cite{Steam}, Amazon-Book \cite{Amazon-book}) is used for initialization. 
% \slh{
% We utilize dataset alignment and LLM-generation methods \cite{Survey_Renmin} to initialize user profiles.}
Each agent's profile contains two components: social traits and unique tastes \cite{social_traits, Social_role}.

Social traits encompass three key traits capturing the element of an individual's personality and characteristics in recommendation scenarios, that is activity, conformity, and diversity.
Activity quantifies the frequency and breadth of a user's interactions with recommended items, delineating between users who extensively watch and rate a number of items and those who confine themselves to a minimal set \cite{proflile_survey}.
The activity trait for user $u$ can be mathematically articulated as: $T_{act}^{u} \doteq \sum_{i\in\Set{I}}y_{ui}$.
Conformity delves into how closely a user's ratings align with average item ratings, drawing a distinction between users with unique perspectives and those whose opinions closely mirror popular sentiments \cite{bias_debias, DICE}.
For user $u$, the conformity trait is defined as: $T_{conf}^u \doteq \frac{1}{\sum_{i \in \Set{I}}y_{ui}}\sum_{i\in\Set{I}}y_{ui} \cdot |r_{ui}-R_i|^2$.
Diversity reflects the user's proclivity toward a diverse range of item genres or their inclination toward specific genres \cite{novelRec}.
The diversity trait for user $u$ is formulated as: $T_{div}^u \doteq |\cup_{i \in \{y_{ui}=1\}}G_i|$.
% Users typically exhibit a long-tailed distribution concerning these social traits.
\slh{Users typically exhibit specific distributions among these social traits (e.g., the long-tail distribution of user activity \cite{user_long_tail}).}
Accordingly, we segment them into three uneven tiers based on each respective trait. 
% For a detailed exposition of the data distribution and the design of prompts across these tiers, refer to Appendix \ref{sec:app_social_traits}. 

To encode users' personalized preferences in natural language, we randomly select 25 items for each user from their viewing history. 
Items rated 3 or above are categorized as `like' by the user, while those rated below 3 are deemed `dislike'. 
Leveraging ChatGPT, we then distill and summarize the unique tastes and rating patterns the user exhibited.
These personalized item tastes are incorporated as the second component into user profiles.

We underscore that in Agent4Rec, certain personal identifiers, such as name, gender, age, and occupation, are intentionally obscured to guarantee widespread applicability and address privacy concerns \cite{demographic_bias}. 
Although these attributes may be instrumental in shaping other types of agents, within the realm of recommendation, they do not dominate users' item preferences. 
Such preferences can be adequately deduced from historical viewing records, rating patterns, and insights embedded in user interactions \cite{private_trait}.

\subsubsection{\textbf{Memory Module.}}
Humans retain diverse memories, bifurcating mainly into factual and emotional categories. 
Of these, emotional memories constitute the core of personal history and exert a stronger influence on decision-making \cite{emotional_memory}.
Although pioneering studies on agents have detailed the architecture of memory, providing foundational blueprints for subsequent explorations, the emotional memories have been largely overlooked \cite{Generative_Agent, RecAgent}.

In our Agent4Rec, we embed a specialized memory module within each generative agent, logging both factual and emotional memories.
Tailored for the recommendation task, factual memories encapsulate interactive behaviors within the recommender system, while emotional memories capture the psychological feeling stemming from these interactions.
Specifically, factual memory mainly contains the list of recommended items, along with user feedback. 
The feedback covers aspects like whether the user watches the item, his corresponding ratings, and potential exit behaviors.
The emotional memory, on the other hand, records user feelings during system interactions, such as levels of fatigue and overall satisfaction.
We aim to ensure the generative agent does not merely react based on past factual interactions but also takes into account the feelings, thereby mirroring genuine human behaviors more closely.

We store memories in two formats: natural language descriptions and vector representations. 
The former is designed for easy understanding by humans \cite{Reflection}. 
While vector representations are primed for efficient memory retrieval and extraction \cite{AutoGen}.

To help agents interact with the recommendation environment, we introduce three memory operations: memory retrieval, memory writing, and memory reflection.
\begin{itemize}[leftmargin = *]
    \item Memory Retrieval: Grounded in insights from studies \cite{Socially_alignment, GITM}, this operation assists the agent in distilling the most relevant information from its memory module.
    \item Memory Writing: This operation enables the recording of agent-simulated interactions and emotions into the memory stream.
    \item Memory Reflection: Recognizing the influence of emotions over user behaviors in recommendations, we incorporate an emotion-driven self-reflection mechanism. 
    This stands in contrast to conventional agent memory designs, such as self-summarization \cite{self-sum}, self-verification \cite{Toolformer}, and self-correction \cite{AoT}, which predominantly condense or deduce advanced factual knowledge, often sidelining emotional feelings. 
    Here, once the agent's actions surpass a pre-defined count, it triggers a reflection process.
    Armed with the prowess of the LLM, the agent introspects its satisfaction with the recommendations and assesses its fatigue levels, offering a deeper understanding of its cognitive state.
\end{itemize}

\subsubsection{\textbf{Action Module.}}
Equipping agents with user profiles and memory modules enables them to exhibit diverse actions akin to humans based on current observations \cite{Survey_Fudan}. 
In Agent4Rec, we design an action module specifically tailored for recommendation domain, which encompasses two broad categories of actions:

\begin{itemize}[leftmargin=*]
    \item Taste-driven Actions: view, rate, and generate post-viewing feelings for items.
    In Agent4Rec, the recommended items are first generated by recommendation algorithms and then presented to agents in a page-by-page manner (further details are available in Section \ref{sec:method_environment}).
    Guided by their taste, agents assess each item on the page for consistency with their preferences. 
    They may choose to watch certain items that pique their interest while bypassing others and subsequently providing ratings and feelings for each item they watch. 
    % Throughout this process, agents may generate some emotional feelings, which are then stored in memory.  
    \item Emotion-driven Actions:
    exit and rate recommender systems, and do post-exit interviews.
    Emotions in the recommendation environment can shape an agent's experience significantly, which are often overlooked in simulations \cite{RecAgent, RecoGym}. 
    An agent's satisfaction with previously recommended items and its current fatigue level influences its decision to continue exploring further recommendation pages or to exit the recommender system. 
    To better simulate this multifaceted decision-making, we enhance the agent's ability for emotional reasoning via Chain-of-Thought \cite{CoT}.
    Initially, the agent discerns the current recommendation page and retrieves its satisfaction level with preceding recommendations from its emotional memory.
    Following this, the agent autonomously expresses its satisfaction and fatigue level for the current recommendation page. 
    Drawing upon these insights, combined with its personalized activity trait, the agent decides whether to exit the system.
    Post-exit, we conduct an interview with each agent, aiming at capturing agents' ratings and overall impressions of the recommender system, offering explicit explanations of their behaviors while navigating the system.
    This interview-style feedback provides a richer and more human-understandable evaluation of the system, enhancing the insights from traditional metrics. 
    An in-depth exploration of the interview can be found in Section \ref{sec:interview}.
\end{itemize}

\subsection{Recommendation Environment} \label{sec:method_environment}
Agent4Rec simulates the interactions between agents and the recommendation environment. 
We discuss three aspects of environment construction that resonate with real-world scenarios, including item profile generation, page-by-page recommendation scenarios, and recommendation algorithm designs.

\begin{itemize}[leftmargin=*]
    \item Item Profile Generation: We construct item profiles to capture key item features, including quality, popularity, genre, and summary.
    Quality is deduced from historical ratings, popularity is based on the number of reviews, while genre and summary are generated by LLM.
    Our goal goes beyond encapsulating the uniqueness of the item in a profile to simulate the recommendation scene of real users.
    We also aim to test whether LLM has potential hallucinations regarding the item. 
    Our approach utilizes a few-shot learning approach, tasking the LLM with classifying the item into one of 18 genres and generating a summary using only the item title. 
    If the LLM's genre classification aligns with the data, it signifies its knowledge of the item.
    To maintain reliability, items causing genre misclassification by the LLM are pruned, reducing hallucination risks. 
    This approach ensures the agent's trustworthiness in simulating user behavior. 
    % More details can be found in Appendix \ref{sec:app_environment}.
    \item Page-by-Page Recommendation Scenario: Our simulator mirrors the operation of real-world recommendation platforms like Netflix, YouTube, and Douban, functioning in a page-by-page manner. 
    Users are initially presented with a list of item recommendations on each page. 
    Based on interactions, preferences, and feedback, subsequent pages could be set to tailor the recommendations further, aiming for a more refined user experience. 
    For further details and experimental results, please refer to Section \ref{sec:exp_aug}. 
    \item Recommendation Algorithm Designs: In Agent4Rec, the recommendation algorithm is structured as a standalone module, with a core focus on extensibility. 
    This design encompasses pre-implemented collaborative filtering-based strategies, including random, most popular, Matrix Factorization (MF) \cite{MF}, LightGCN \cite{LightGCN}, and MultVAE \cite{MultVAE}. 
    Moreover, it incorporates an open interface, enabling researchers and practitioners to effortlessly integrate external recommendation algorithms. 
    This adaptability ensures that Agent4Rec could be a versatile platform for comprehensive evaluations and the collection of valuable user feedback in the future.
\end{itemize}
 
\section{Agent Alignment Evaluation.}
With the specialized simulator, Agent4Rec, tailored for recommendations, we would like to explore an essential research question: 
\begin{itemize}[leftmargin=*]
    \item \textbf{RQ1:} To what extent can LLM-empowered generative agents truly simulate the behaviors of genuine, independent humans in recommender systems?
\end{itemize}
% Before advancing toward a sufficiently realistic simulation platform for recommender systems, we need to answer the following question: To what extent can LLM-driven agents mimic human user behaviors in the real world? 
In this section, we will delve into the capability and limitations of generative agents from various perspectives, including the alignment of user behavior (such as user taste, rating distribution, and social traits) and the evaluation of the recommendation environment (including recommendation \slh{strategies evaluation}, page-by-page recommendation enhancements, and the case study of interview).

\begin{table*}[t]
    \centering
    \caption{User Taste Alignment across Movielens, Amazon-Book, and Steam datasets}
    \label{tab:discrimination}
    \vspace{-10pt}
    \resizebox{0.9\linewidth}{!}{
    \begin{tabular}{c|cccc|cccc|cccc}
    \toprule
    \multicolumn{1}{c|}{} & \multicolumn{4}{c|}{\textbf{MovieLens}} & \multicolumn{4}{c|}{\textbf{Amazon-Book}} & \multicolumn{4}{c}{\textbf{Steam}}  \\
        1:m & Accuracy & Recall & Precision & F1 Score & Accuracy & Recall & Precision & F1 Score & Accuracy & Recall & Precision & F1 Score \\ \midrule
        1:1 & \textbf{0.6912*} & 0.7460 & \textbf{0.6914*} & \textbf{0.6982*} & \textbf{0.7190*} & \textbf{0.7276*} & \textbf{0.7335*} & \textbf{0.7002*} & \textbf{0.6892*} & 0.7059 & \textbf{0.7031*} & \textbf{0.6786*} \\ 
        1:2 & 0.6466 & 0.7602 & 0.5058 & 0.5874 & 0.6842 & 0.6888 & 0.5763 & 0.5850 & 0.6755 & 0.7316 & 0.5371 & 0.5950 \\ 
        1:3 & 0.6675 & 0.7623 & 0.4562 & 0.5433 & 0.6707 & 0.6909 & 0.4423 & 0.5098 & 0.6505 & \textbf{0.7381*} & 0.4446 & 0.5194 \\ 
        1:9 & 0.6175 & \textbf{0.7753*} & 0.2139 & 0.3232 & 0.6617 & 0.6939 & 0.2369 & 0.3183 & 0.6021 & 0.7213 & 0.1901 & 0.2822 \\ 
    \bottomrule
    \end{tabular}}
    \vspace{-3pt}
\end{table*}

\smallskip\noindent\textbf{Motivation.} 
To appropriately respond to recommended items, generative agents need to have a clear understanding of their own preferences.
We conjecture that an independent, personalized agent, initialized from real users in MovieLens-1M, should maintain long-term preference coherence.
In practice, this implies that the agent should be adept at distinguishing the items that real users favor.

\smallskip\noindent\textbf{Setting.}
% We ask generative agents to discriminate between the items the corresponding real users have watched and those they have not, to validate the authenticity of the tastes recorded in the user profile module.
To validate how well generative agents align with the preferences encoded in their user profiles, we task agents with distinguishing between the items that the corresponding real users have interacted with and those they have not. \slh{We conduct experiments on three real-world datasets (\ie Movielens-1M \cite{MovieLens}, Steam\cite{Steam}, Amazon-Book\cite{Amazon-book}).}
Specifically, a total of 1,000 agents will each be randomly assigned 20 items.
Among these, the ratio between items the user has interacted with (\ie $y_{ui} = 1$) but was not utilized for profile initialization and items the user has not interacted with (\ie $y_{ui} = 0$) is set as $1:m$, with $m\in\{1,2,3,9\}$. 
% In particular, for each item that a user has watched (\ie $y_{ui} = 1$) but was not utilized for profile initialization, we randomly sample $m \in\{1,2,3\}$ items that the user has not watched (\ie $y_{ui} = 0$).
% we ask the user to decide whether to watch these $(1+m)k$ items based on their preferences, where $k$ is a scaling factor.
Under this setting, agent responses (\ie $\hat{y}_{ui}$) to recommended items are considered binary discrimination, taking values between 0 and 1.

\smallskip\noindent\textbf{Results.} 
Table \ref{tab:discrimination} reports the empirical discrimination results across various metrics.
The best performance for each metric is highlighted in bold and marked with an asterisk. 
We observe that:
\begin{itemize}[leftmargin=*]
    \item \textbf{Generative agents consistently and impressively identify items aligned with user preferences.}
    Specifically, regardless of the number of distracting items introduced, agents maintain a high accuracy of around 65\% and recall of about 75\%. 
    This high fidelity can be attributed to the personalized profile faithfully mirroring the user's genuine interests. 
    It further reflects that the agents effectively encapsulate a substantial portion of the real preferences, signifying the feasibility of LLM-powered generative agents in recommendation simulation.
    \item \textbf{Agents tend to maintain a relatively consistent count of preferred items, a phenomenon potentially stemming from inherent hallucinations in the LLM.}
    Notably, while Accuracy and Recall show satisfactory results, metrics like Precision and F1 Score, which emphasize the cost of false positives, experience a sharp decline from nearly 70\% to about one quarter as the proportion of user-liked items decreases. 
    We attribute this failure to LLM's inherent hallucinations that agents tend to consistently pick a set number of items.
    Such behavior highlights challenges in designing a more reliable recommendation simulator using LLM-empowered generative agents. 
    However, we emphasize that in the subsequent simulation results with recommendation algorithms, a substantial proportion of recommended items align with user preferences, thereby endorsing high trustworthiness in those simulation outcomes.
\end{itemize}
% We compute the typical metrics for binary classification in Table \ref{tab:discrimination}.
% Under the $m=1$ setting, we achieve an accuracy of $69.12\%$ while consistently sustaining a recall score of approximately $75\%$ across all settings. This suggests that generative agents effectively capture the majority of their genuine preferences.

% \subsubsection{Rating Distribution Alignment}
\subsection{Rating Distribution Alignment}

\begin{figure}[t]
	\centering
	\subcaptionbox{Distribution on MovieLens \label{fig:rating_true}}{
	    \vspace{-5pt}
	\includegraphics[width=0.47\linewidth]{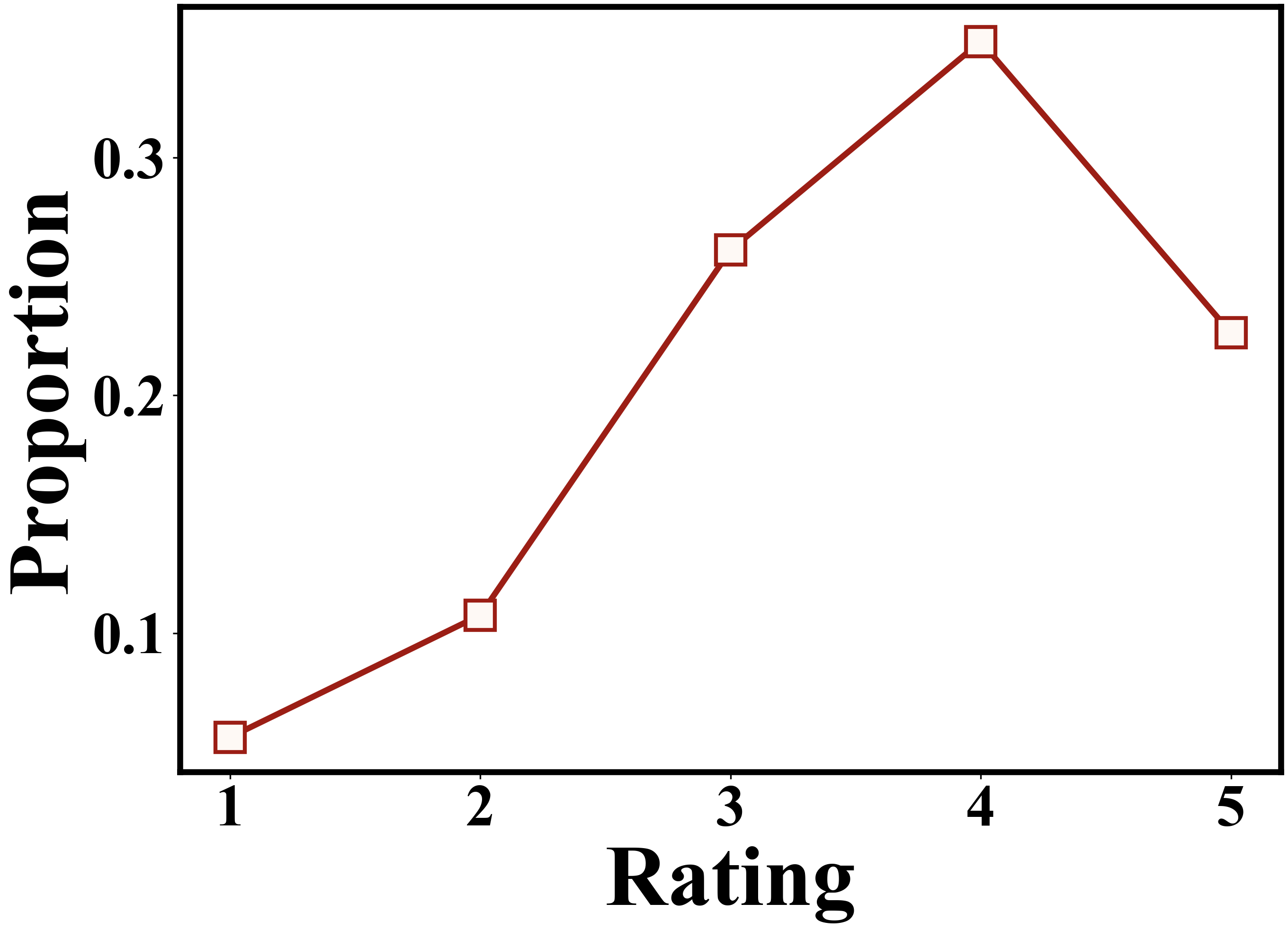}}
	\subcaptionbox{Agent-simulated distribution\label{fig:rating_sim}}{
	     \vspace{-5pt}
	\includegraphics[width=0.47\linewidth]{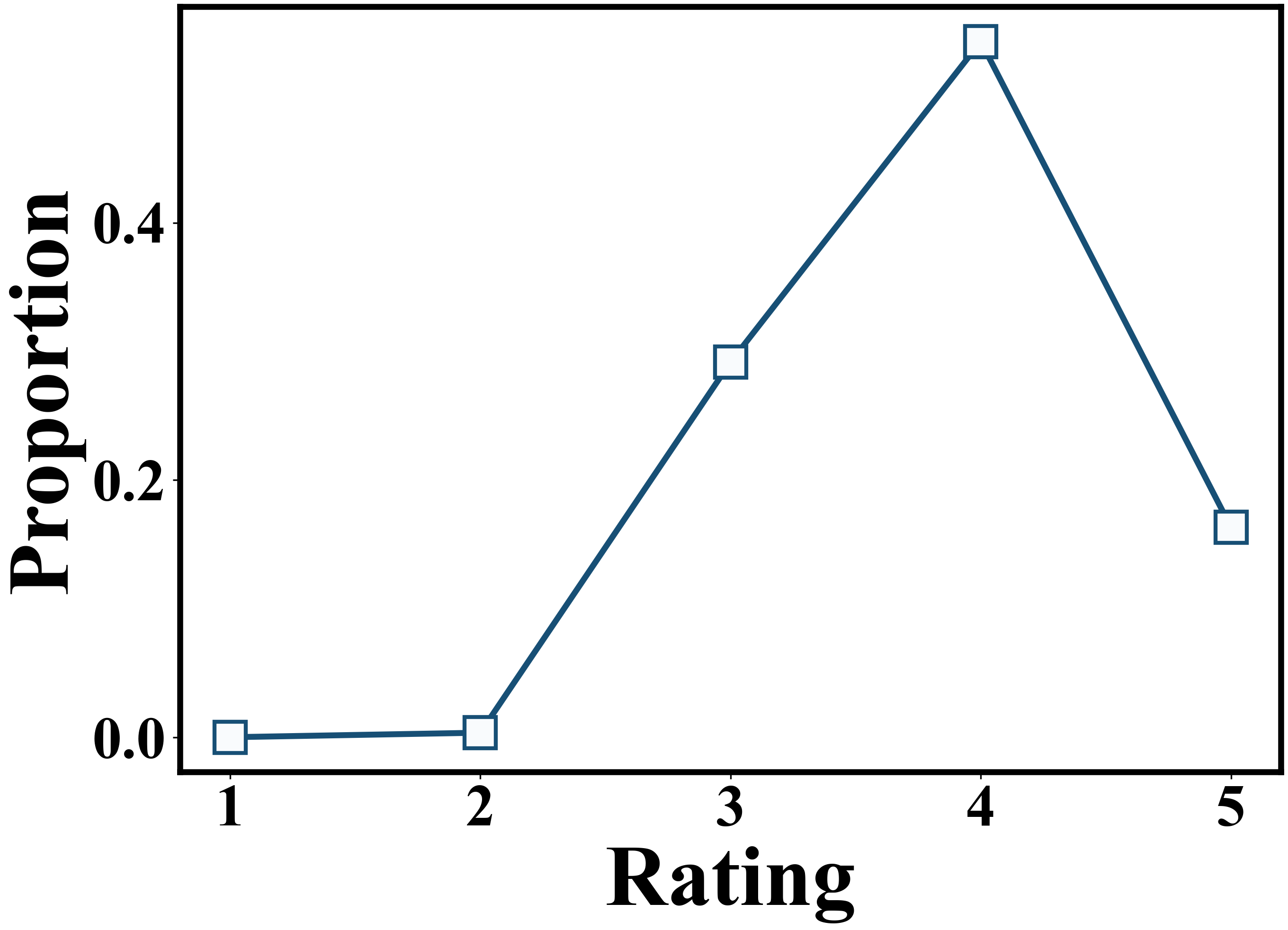}}
	\vspace{-10pt}	
	\caption{Comparison between ground-truth and agent-simulated rating distributions.}
	\label{fig:rating_dist}
	% \vspace{-5pt}
\end{figure}
\textbf{Motivation.} Beyond ensuring user-aligned behaviors at a micro-level for each agent, a comprehensive evaluation of the simulator requires that the generative agents accurately mirror real-world user behavioral patterns at a macro scale. 
Specifically, in terms of rating distribution, the goal is to maintain consistency with MovieLens-1M data distributions.

\smallskip\noindent\textbf{Results.} 
Figures (\ref{fig:rating_true}) and (\ref{fig:rating_sim}) illustrate the ground-truth rating distribution in the Movielens-1M dataset and the distribution of ratings generated by agents, respectively. 
The figures demonstrate a strong alignment between the simulated rating distribution and the actual distribution. 
Specifically, ratings at 4 dominate the overall distribution, while low ratings (1-2) constitute only a small portion. 
It is worth mentioning that agents tend to give very few 1-2 ratings, which differs from genuine human behavior.
This is attributed to the LLM's extensive prior knowledge of movies, as agents tend to avoid watching low-quality films in advance.
Consequently, simulating the user action of giving low ratings after choosing to watch a movie becomes challenging. 
% This is due to their vast prior knowledge of movies, LLM-based agent tends to avoid watching low-quality movies in advance, making it challenging to authentically simulate the behavior of giving low ratings after choosing to watch a film.
Moreover, Figure (\ref{fig:rating_sim}) represents the simulation results under the MF algorithm, and similar trends are observed under other recommendation algorithms. 
This consistency across different algorithms validates the potential of Agent4Rec.
We demonstrate through ablation that the social traits component of our agents contribute critically to the believability of agent behavior.

\subsection{Social Traits Alignment}
\begin{figure}[t]
	\centering
	\subcaptionbox{Activity\label{fig:bar_sim_act}}{
	    \vspace{-5pt}
		\includegraphics[width=0.31\linewidth]{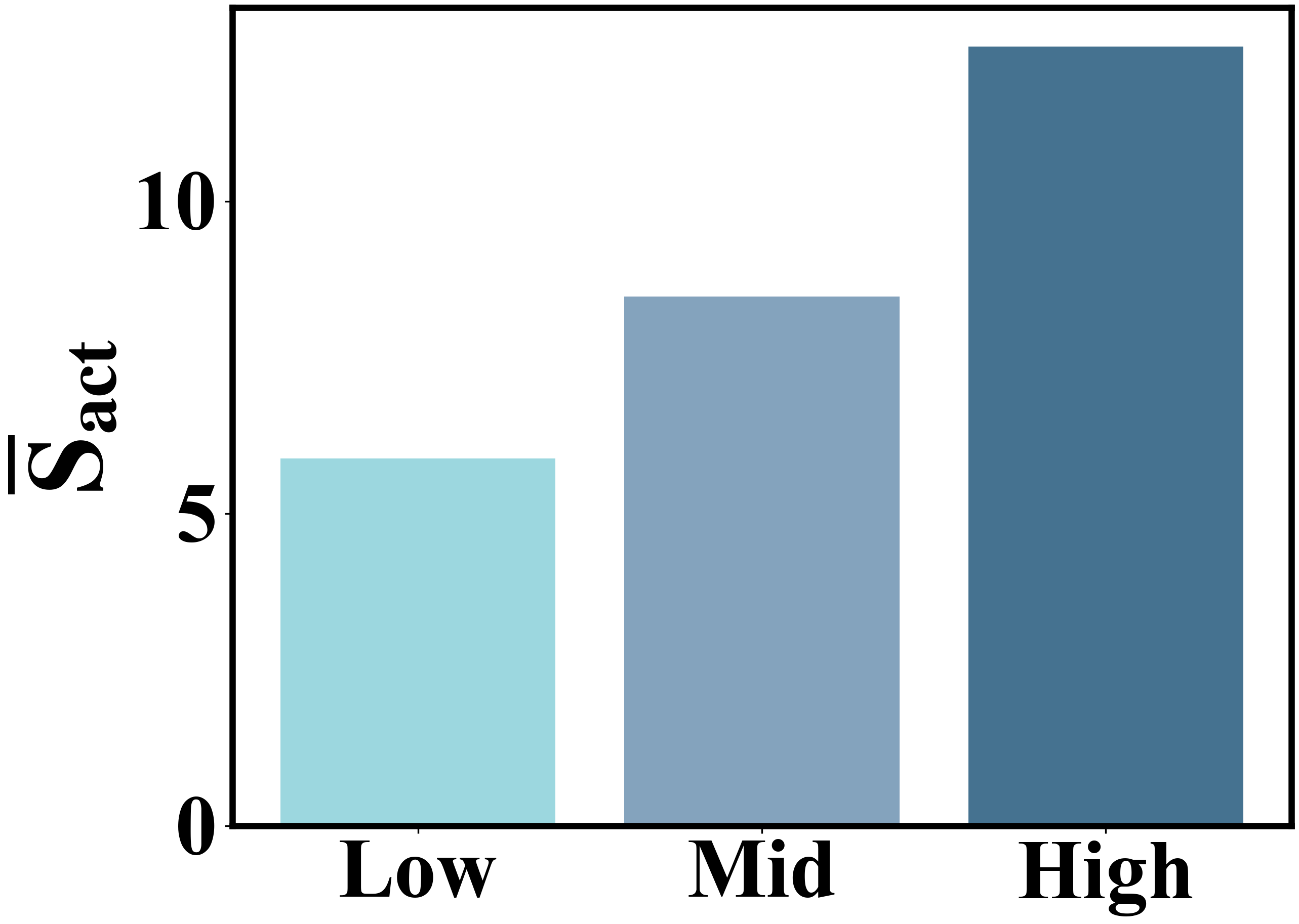}}
	\subcaptionbox{Conformity\label{fig:bar_sim_conf}}{
	    \vspace{-5pt}
		\includegraphics[width=0.31\linewidth]{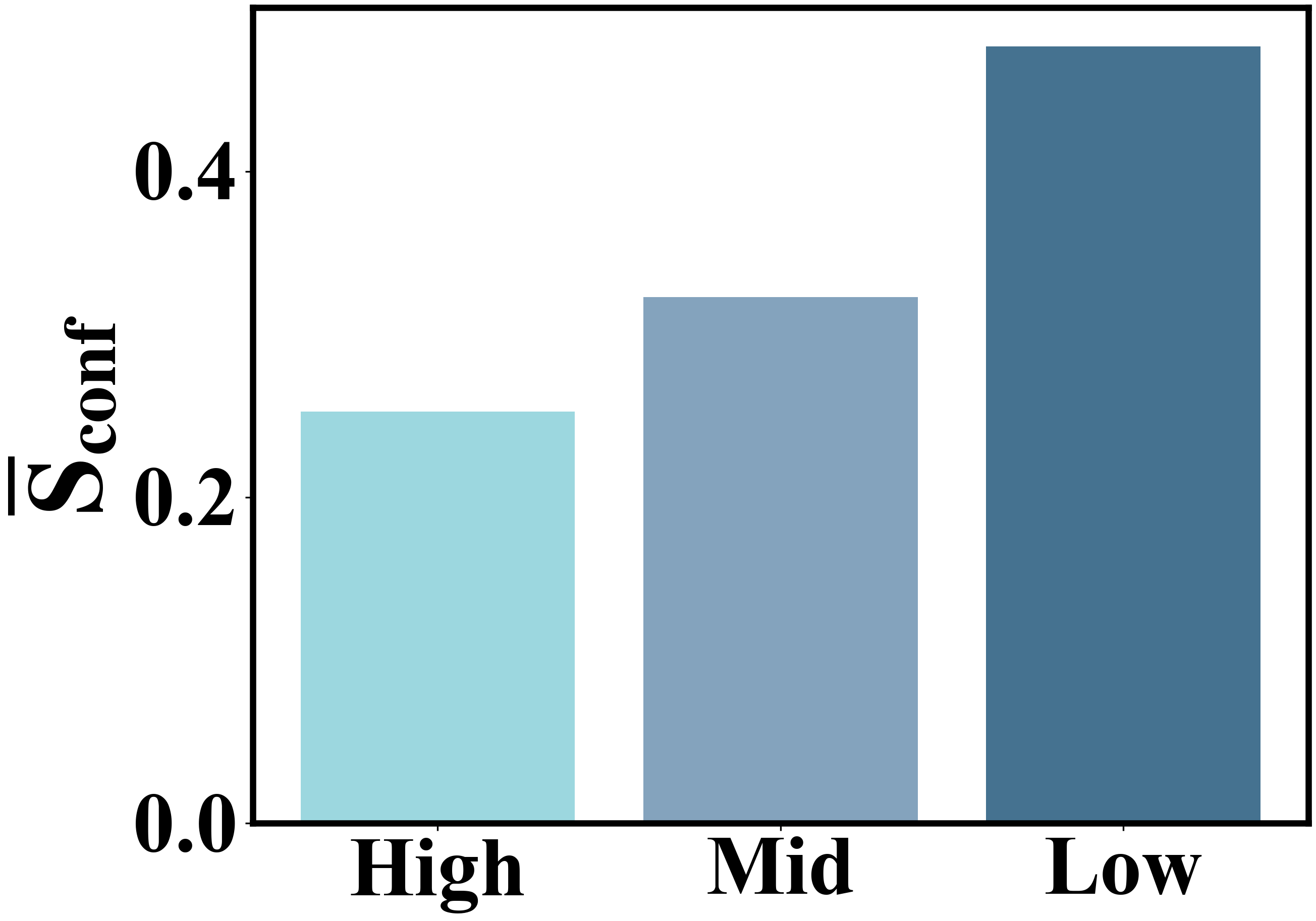}}
	\subcaptionbox{Diversity\label{fig:bar_sim_diver}}{
	    \vspace{-5pt}
		\includegraphics[width=0.31\linewidth]{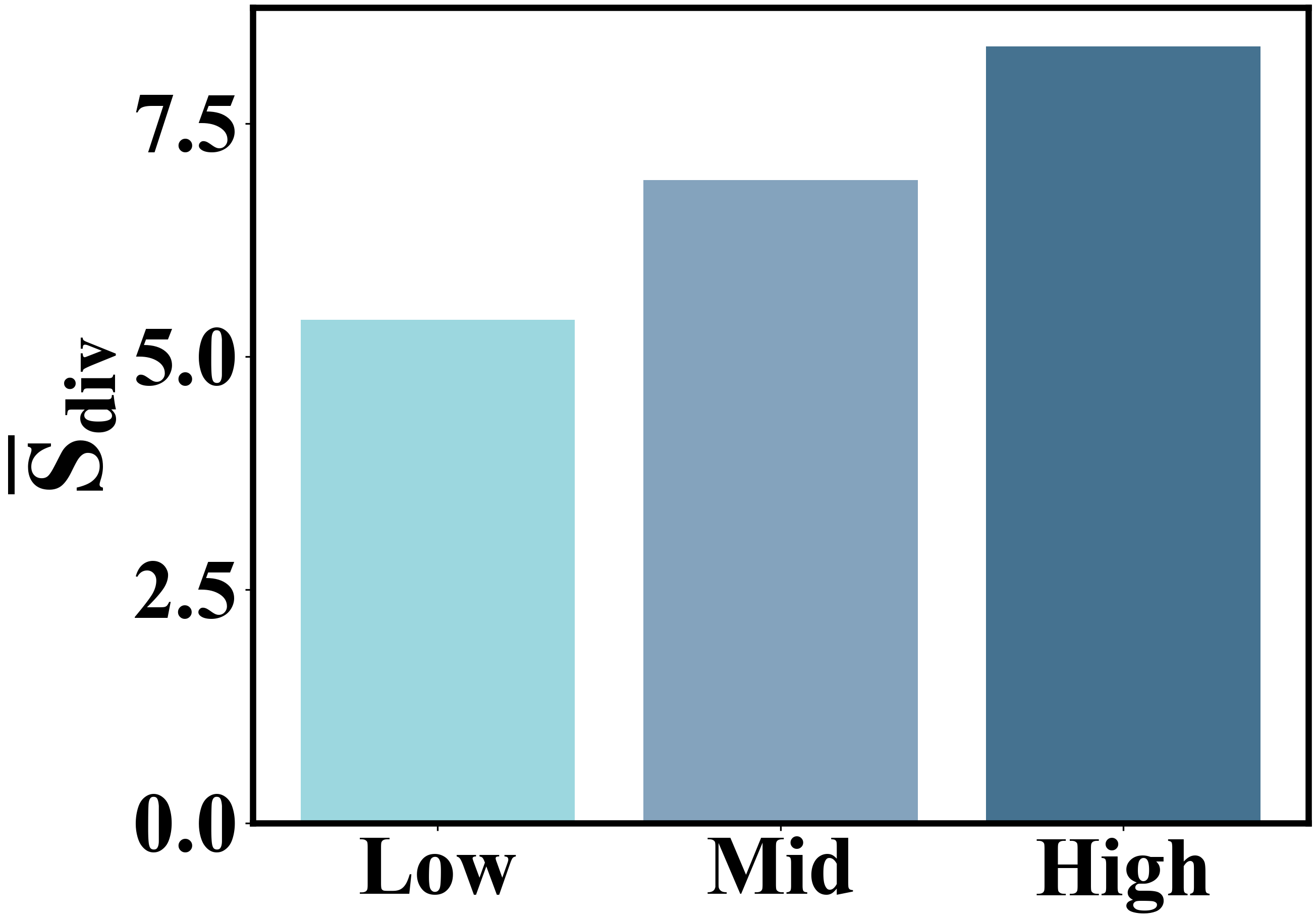}}
	\vspace{-10pt}	
	\caption{Averaged scores of activity, conformity, and diversity among agent groups with varying degrees of social traits.}
        % \caption{\slh{The averaged social traits of agents within different traits groups}.}
	\label{fig:group_sim}
	\vspace{-10pt}
\end{figure}
\textbf{Motivation.} 
In real-world recommendation scenarios, user behavior patterns are influenced by various factors, not solely determined by personal preferences but also impacted by social attributes such as activity, conformity, and diversity \cite{proflile_survey}. 
Accordingly, we specially designed a user profile for Agent4Rec that incorporates these social traits. 
Following the definitions in Section \ref{sec:method_profile} and based on statistics from MovieLens, agents are categorized into three different levels (high, medium, and low) for each trait, giving a consistent prompt for agents within each tier. 
We conjecture that agents exhibiting similar preferences might still display differentiated behavior patterns based on their unique social traits.

\begin{figure}[t]
	\centering
	\subcaptionbox{w Activity trait\label{fig:act_hist}}{
	    \vspace{-5pt}
		\includegraphics[width=0.47\linewidth]{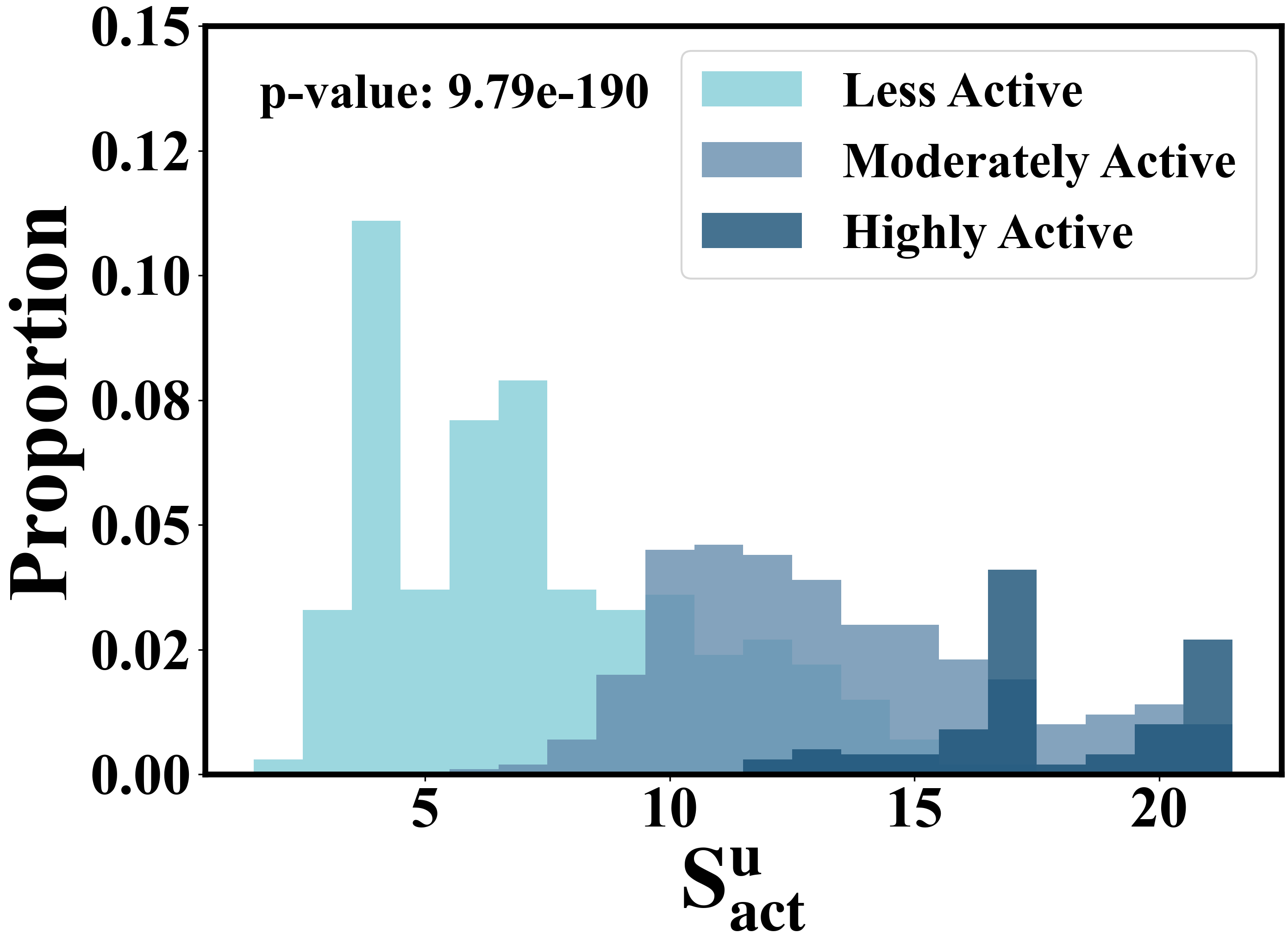}}
	\subcaptionbox{w/o Activity trait\label{fig:act_hist_abl}}{
	    \vspace{-5pt}
		\includegraphics[width=0.47\linewidth]{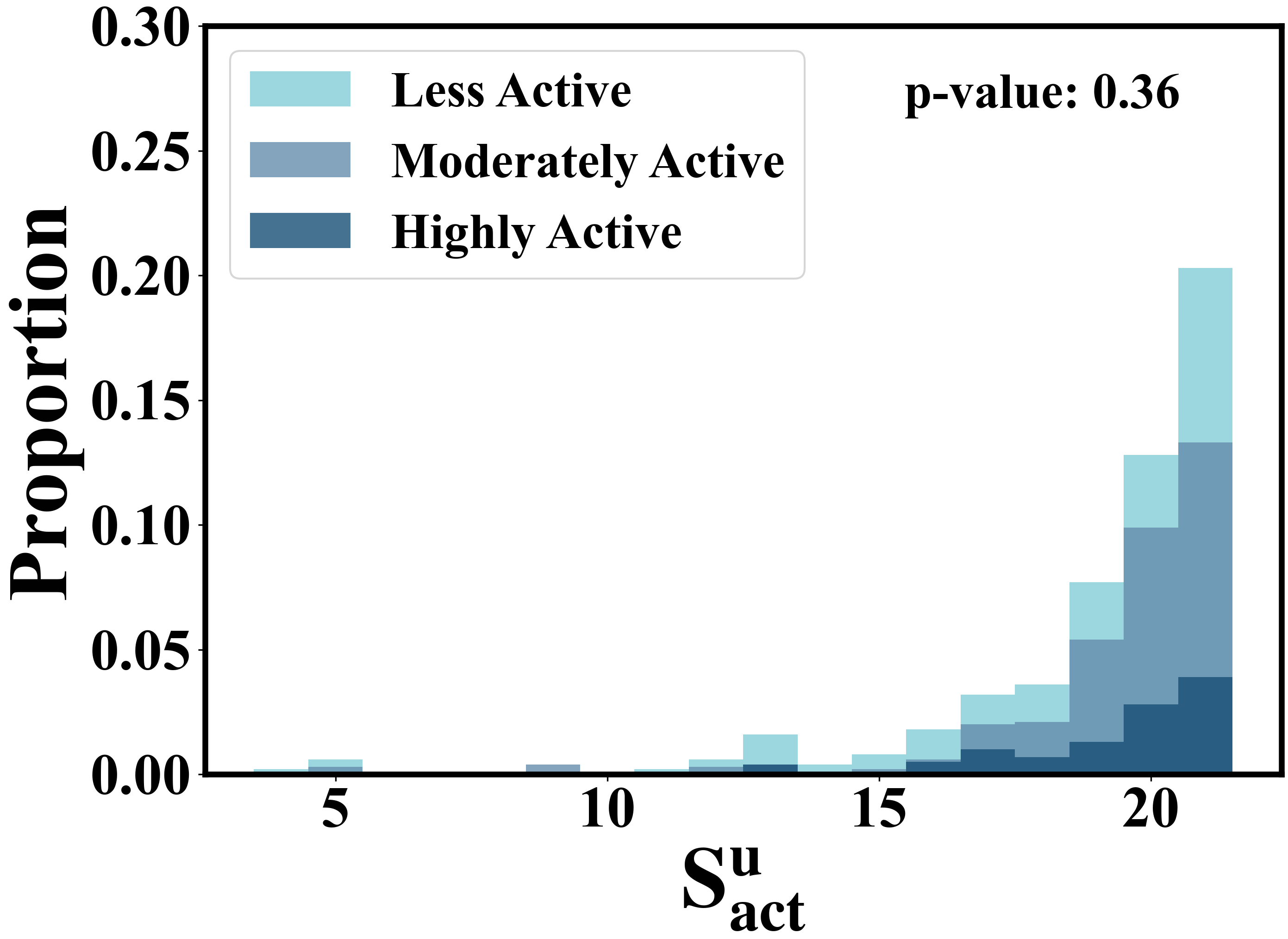}}
	\vspace{-10pt}	
	\caption{Distribution of interaction numbers among agents with varying levels of user activity.}
	\label{fig:hist_act}
	% \vspace{-5pt}
\end{figure}

\begin{figure}[t]
	\centering
	\subcaptionbox{Ground-truth diversity\label{fig:true_diver_rolling}}{
	    \vspace{-5pt}
		\includegraphics[width=0.47\linewidth]{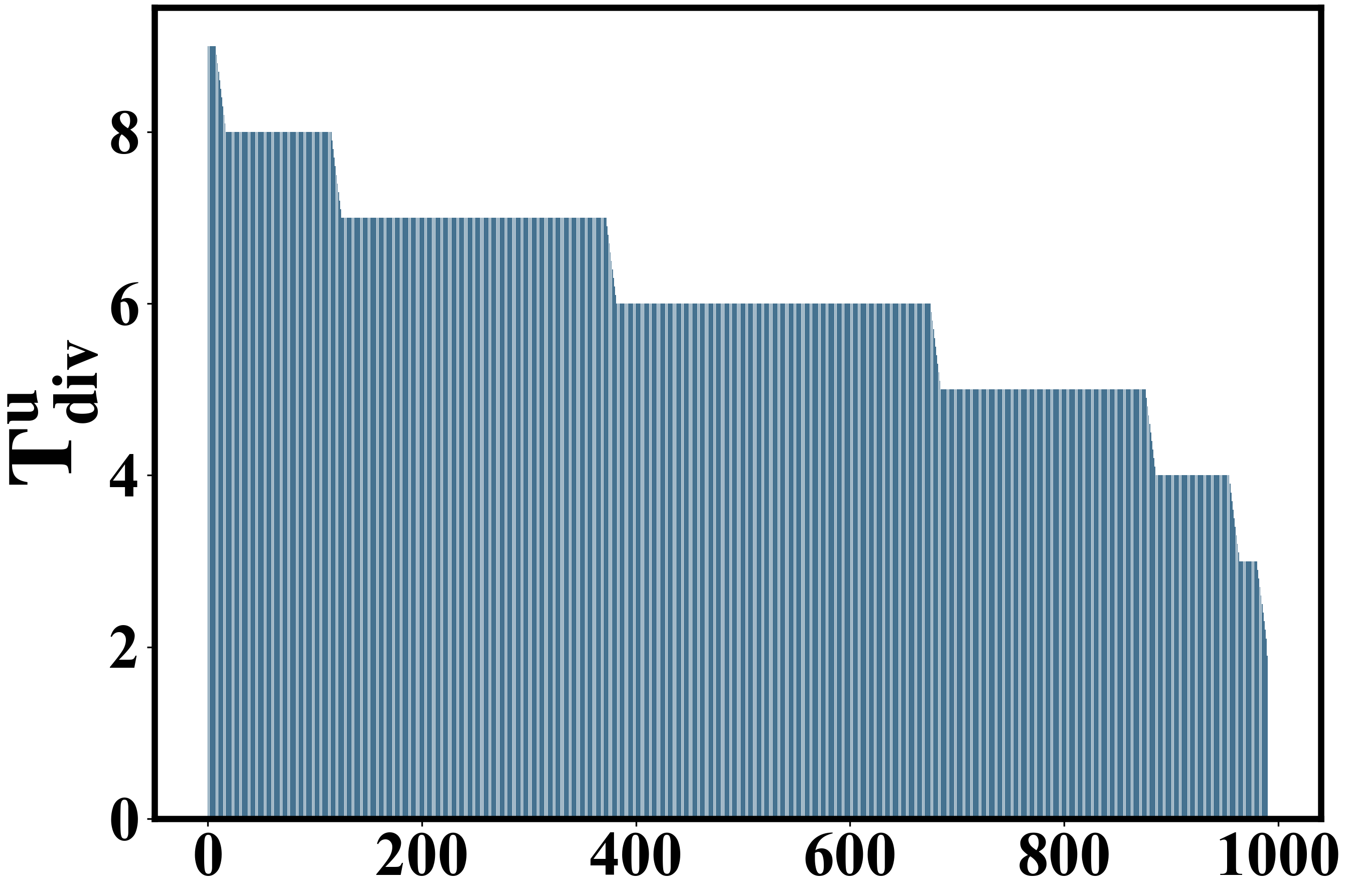}}
	\subcaptionbox{Simulated diversity\label{fig:sim_diver_rolling}}{
	    \vspace{-5pt}
		\includegraphics[width=0.48\linewidth]{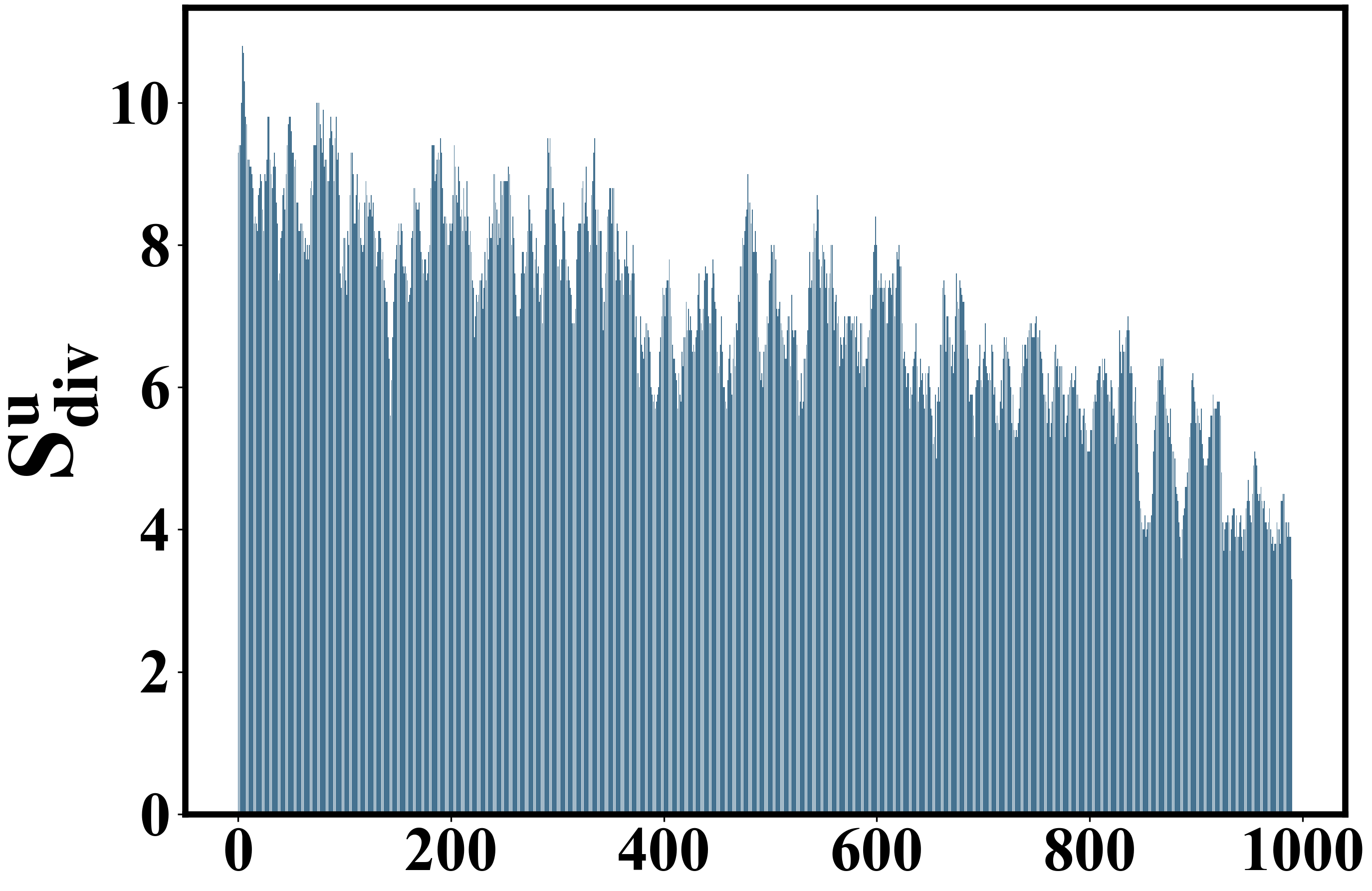}}
	\vspace{-10pt}	
	\caption{Comparison between the \slh{individual level} distributions of ground-truth and agent-simulated diversity scores.}
	\label{fig:rolling_diver}
	\vspace{-10pt}
\end{figure}

\smallskip\noindent\textbf{Results.} To validate the significance of the user profile module's design, it is essential to probe from multiple perspectives, including (1) whether agents categorized into different tiers exhibit distinct behaviors (Figure \ref{fig:group_sim}); (2) whether ablation studies can detect differences between agents with and without social traits in their profiles (Figure \ref{fig:hist_act}); and (3) whether \slh{the simulated traits distribution of agents aligns with that of actual users} (Figure \ref{fig:rolling_diver}).

% For the \slh{detailed} variables introduction, experiment settings, and results in this section, please refer to Appendix \ref{sec:traits_appendix}.
Extensive experimental results from distribution analysis, ablation studies, and statistical tests demonstrate the pivotal role of social traits in user profile construction, significantly influencing agent behavior. 
This further illustrates the universal applicability and referential value of Agent4Rec's setup, which is specifically designed for recommendation tasks. 
It's worth noting that the diversity trait might exhibit minimal distinction in agent behavior, possibly due to the strong overlap of movie categories in the MovieLens dataset, suggesting further research and validation with alternative datasets.

% Figure \ref{fig:act_hist} illustrates that, with social traits prompting, the interactions of generative agents exhibit a long-tail distribution. 
% In contrast, Figure \ref{fig:act_hist_abl} indicates that generative agents exhibit a consistent tendency in interaction proactiveness without such control, deviating from real-world scenarios.
% We also conduct similar experiments on other social traits, namely conformity and diversity. See more details in the Appendix \ref{sec:traits_appendix}.

% \begin{figure}[t]
%     \vspace{10pt}
%     \includegraphics[width=1\columnwidth]{figures/alignment/bar_true.png}
%     \vspace{-20pt}
%     \caption{Group Behaviors of Humans}
%     \vspace{-10pt}
%     \label{fig:bar_true}
% \end{figure}

% \begin{figure}[t]
%     \vspace{10pt}
%     \includegraphics[width=1\columnwidth]{figures/alignment/bar_sim.png}
%     \vspace{-20pt}
%     \caption{Group Behaviors of Agents}
%     \vspace{-10pt}
%     \label{fig:bar_true}
% \end{figure}

\subsection{Recommendation Strategy Evaluation}\label{sec:rec_eval}
\textbf{Motivation.} Human users display varying levels of satisfaction with different recommendation Algorithms. 
For example, users generally gain higher satisfaction with advanced strategies, compared to random recommendations. 
Assuming generative agents can accurately simulate genuine human behaviors, it's plausible that these agents would exhibit similar satisfaction trends to those observations in humans.

\smallskip\noindent\textbf{Setting.} 
For a fair comparison, we employ mainly collaborative filtering-based recommendation strategies (\ie random, most popular, MF \cite{MF}, LightGCN \cite{LightGCN}, and MultVAE \cite{MultVAE}) in our simulator and evaluate their satisfaction trends on MovieLens. 
In Agent4Rec, four movies are displayed on each recommendation page, and agents will take actions such as viewing and rating based on their personal preferences. 
Simultaneously, agents decide whether to proceed to the next recommended page or exit the recommendation system based on their satisfaction. 
The system enforces agents' exit after browsing a maximum of five pages. 
Once an agent exits, we request him to give a satisfaction score for the recommendation system, ranging from 1 to 10. 
Regarding rating higher than 3 as a signal of like, we collect the following multi-facets metrics after the whole simulation completes: average viewing ratio (\ie $\overline{P}_{view}$), the average number of likes (\ie $\overline{N}_{like}$), the average ratio of likes (\ie $\overline{P}_{like}$), the average number of exit page (\ie $\overline{N}_{exit}$), and the average user satisfaction score (\ie $\overline{S}_{sat}$). 
% The detailed calculation for these metrics can be found in Appendix \ref{sec:agorithms_appendix}.

\smallskip\noindent\textbf{Results.}
Table \ref{tab:algorithms} reports the multi-facet satisfaction metrics for various recommendation strategies.
It is clear that agents exhibit higher satisfaction with algorithm-based recommendations compared to random and popularity-based recommendations.
% \textbf{Users exhibit higher satisfaction with algorithm-based recommendation systems compared to random and popularity-based recommendations.} 
This phenomenon aligns with observations in the real world, where well-designed recommendation algorithms can effectively address the issue of information overload in modern society, thereby enhancing users' online experience \cite{rec_survey}.
Furthermore, in line with prevailing insights in the research community, LightGCN outperforms both MF and MultVAE across diverse evaluation criteria. 
Such findings underscore the fine-grained evaluation capabilities of LLM-empowered agents for recommendation strategies. 
This further shed light on the potential of an agent-driven recommendation simulator for A/B testing, offering a more cost-efficient alternative to conventional online A/B testing.
% Moreover, consistent with fundamental knowledge in the research community, LightGCN demonstrates an overall superior performance over MF and MultVAE across multiple evaluative metrics.
% \textbf{Within the algorithm-based methods, LightGCN demonstrates an overall superior performance.} 
% Based on the above observations, we find that LLM-empowered agents exhibit fine-grained distinctions ability for recommendation strategies. 
% This shed light on the potential of conducting an A/B test on a well-designed LLM-agent-based recommendation simulator, thereby reducing the substantial costs for online A/B testing.

\begin{table}[t]
    \centering
    % \vspace{-5pt}
    \caption{Recommendation strategies evaluation.}
    \label{tab:algorithms}
    \vspace{-5pt}
    \resizebox{0.8\linewidth}{!}{
    \begin{tabular}{l|ccccc}
    \toprule
    \multicolumn{1}{c|}{} & $\overline{P}_{view}$ & $\overline{N}_{like}$ & $\overline{P}_{like}$ & $\overline{N}_{exit}$ & $\overline{S}_{sat}$
    \\\midrule
     Random   & 0.312 & 3.3 & 0.269 & 2.99 & 2.93 \\
    Pop & 0.398 & 4.45 & 0.360 & 3.01 & 3.42 \\
    MF  & 0.488 & \textbf{6.07*} & 0.462 & \textbf{3.17*} & 3.80 \\
    MultVAE & 0.495 & 5.69 & 0.452 & 3.10 & 3.75 \\
    LightGCN & \textbf{0.502*} & 5.73 & \textbf{0.465*} & 3.02 & \textbf{3.85*}\\\bottomrule
    % Imp.\%    & & & \\\bottomrule
    \end{tabular}}
    \vspace{-10pt}
\end{table}

\subsection{Page-by-Page Recommendation Enhancement}\label{sec:exp_aug}

\begin{table}[t]
    \centering
    % \vspace{-5pt}
    \caption{Page-by-page recommendation enhancement results over various algorithms.}
    \label{tab:augmentation}
    \vspace{-5pt}
    \resizebox{1\linewidth}{!}{
    \begin{tabular}{l|cc|cc|cc}
    \toprule
      \multicolumn{1}{c|}{} & \multicolumn{2}{c|}{MF} & \multicolumn{2}{c|}{MultVAE} & \multicolumn{2}{c}{LightGCN}  \\
    \multicolumn{1}{c|}{Offline} & Recall & \multicolumn{1}{c|}{NDCG} & Recall & \multicolumn{1}{c|}{NDCG} & Recall & \multicolumn{1}{c}{NDCG} 
    \\\midrule
     Origin    & 0.1506 & 0.3561 & 0.1609 & 0.3512 & 0.1757 & 0.3937 \\
     % + Random & 0.1464 & 0.3501 & 0.1599 & \note{0.3580} & 0.1733 & 0.3902 \\
    + Unviewed & 0.1523 & 0.3557 & 0.1598 & 0.3487 & 0.1729 & 0.3849 \\
    + Viewed & \textbf{0.1570*} & \textbf{0.3604*} & \textbf{0.1613*} & \textbf{0.3540*} & \textbf{0.1765*} & \textbf{0.3943*}\\\midrule\midrule
    % \multicolumn{1}{c|}{} & \multicolumn{2}{c|}{MF} & \multicolumn{2}{c|}{MultVAE} & \multicolumn{2}{c}{LightGCN}  \\
    
    \multicolumn{1}{c|}{Simulation} & $\overline{N}_{exit}$ & \multicolumn{1}{c|}{$\overline{S}_{sat}$} & $\overline{N}_{exit}$ & \multicolumn{1}{c|}{$\overline{S}_{sat}$} & $\overline{N}_{exit}$ & \multicolumn{1}{c}{$\overline{S}_{sat}$} 
    \\\midrule
     Origin    & 3.17 & 3.80 & 3.10 & 3.75 & 3.02 & 3.85 \\
     % + Random & 0.1464 & 0.3501 & 0.1599 & \note{0.3580} & 0.1733 & 0.3902 \\
    + Unviewed & 3.03 & 3.77 & 3.01 & 3.77 & 3.06 & 3.81 \\
    + Viewed & \textbf{3.27*} & \textbf{3.83*} & \textbf{3.18*} & \textbf{3.87*} & \textbf{3.10*} & \textbf{3.92*}\\
    \bottomrule
    \end{tabular}}
    \vspace{-10pt}
\end{table}

\textbf{Motivation.} 
In real-world settings, recommendation platforms frequently collect immediate user behaviors to iteratively refine the recommender, aiming to accurately capture users' latest preferences. 
Our aim in designing the page-by-page recommendation setting is to emulate this feedback-driven recommendation enhancement.

\smallskip\noindent\textbf{Setting.} 
After a complete recommendation simulation, we collect both viewed and unviewed movies for each agent. 
These two types of movies are then added as positive signals to the training set of each user to re-train the recommendation algorithms. 
We evaluate the performance of these retrained recommenders using standard offline metrics, such as Recall@20 and NDCG@20, as well as satisfaction simulation evaluations, namely $\overline{N}_{exit}$ and $\overline{S}_{sat}$.

\smallskip\noindent\textbf{Results.} 
As depicted in Table \ref{tab:augmentation}, by leveraging movies \slh{viewed} by the agent as augmented data, all recommendation algorithms exhibit improvements in both offline evaluation metrics and simulated satisfaction evaluations. 
However, when the training dataset is augmented with unviewed movies, the overall user experience typically deteriorates. Successfully emulating this feedback-driven recommendation augmentation indicates that movie choices by the agent can serve as a consistent indicator of a user's unique preferences.
% Table \ref{tab:augmentation} reports both offline and online evaluation metrics for re-trained recommendation algorithms.
% The results show that by adopting movies viewed by agents into the training set, all the recommendation algorithms show improvements in both offline evaluation metrics and satisfaction metrics. In contrast, augmenting the training dataset with unviewed movies generally leads to a decline in user experience. This phenomenon suggests the viewing action of agents aligns with their unique tastes faithfully.

% \subsubsection{Case Study: Feeling and Explanation}
\subsection{Case Study of \slh{Feeling} Interview} \label{sec:interview}
\textbf{Motivation.} 
Compared to conventional recommendation simulators \cite{RecSim, Virtual-Taobao}, the unique strength of LLM-empowered agent simulation lies in its ability to provide human-comprehensible explanations  \cite{InteRecAgent}.
Eliciting explanations from agents provides insights into the reliability of simulations, enabling us to further refine the recommender system.

\smallskip\noindent\textbf{Results.} 
Figure \ref{fig:InterviewCase} presents a case from post-exit interviewing on MovieLens. 
The agent rates the recommended movies based on their personal tastes, social traits, and emotional memory. 
Specifically, the agent recognized that the recommender system did suggest movies in line with their preferences. 
However, there were aspects leading to less satisfaction. 
For instance, even though the agent has diverse interests, the system tended to recommend popular movies.
% Figure \ref{fig:InterviewCase} illustrates one record of interviewing an agent after finishing the simulation. This agent systematically rated the recommendation results considering both personal tastes and social traits. Specifically, the user acknowledged that the recommendation system did recommend movies that aligned with his preferences, but there were aspects that left him less satisfied. For instance, despite he has diverse interests, the system tended to recommend an excessive number of popular items.

\begin{lrbox}{\InterviewCase}
\begin{tcolorbox}[colback=background_u,colframe=frame_u,boxrule=0.5pt,title={\textbf{Case Study of Feeling Interview}}]
\textbf{Question:}
\textit{Do you feel satisfied with the recommender system? Rate it from 1-10 and give an explanation.}\\
% you have just interacted with? Rate this recommender system from 1-10 and give an explanation. \\
\textbf{Rating:} 
\textit{6}\\
\textbf{Reason:}
\textit{While the recommender system did provide me with some movies that aligned with my taste, there were also a few recommendations that I disliked. The system took into account my personal preferences and historical ratings, which I appreciate as a Balanced Evaluator. However, as an Occasional Viewer, I am not easily swayed by movie recommendations and only seek out movies that strictly align with my taste. The system failed to consistently meet my preferences, resulting in a few unsatisfied memories. Additionally, as a Cinematic Trailblazer, I enjoy exploring unique and obscure movies, but the recommendations seemed to focus more on popular classics. Overall, the system had its strengths but fell short of fully satisfying my movie-watching preferences.}
% While the recommender system did provide me with some movies that aligned with my taste, there were also a few recommendations that I disliked. The system took into account my personal preferences and historical ratings, which I appreciate as a Balanced Evaluator. However, as an Occasional Viewer, I am not easily swayed by movie recommendations and only seek out movies that strictly align with my taste. The system failed to consistently meet my preferences, resulting in a few unsatisfied memories. Overall, the system had its strengths but fell short of fully satisfying my movie-watching preferences.
\end{tcolorbox}
\end{lrbox}

\begin{figure}[t]
    \centering
    % \vspace{-10pt}
    \usebox{\InterviewCase}
    \vspace{-15pt}
    \caption{One case study of post-exit feeling interview.}
    \label{fig:InterviewCase}
    \vspace{-15pt}
\end{figure}

\section{Insights and Exploration}

Given Agent4Rec's promising simulation capabilities, we pose another profound research question: 
\begin{itemize}[leftmargin=*]
    \item \textbf{RQ2:} Can Agent4Rec provide some insights on unresolved problems in the recommendation domain? 
\end{itemize}
In this section, we discuss two insights drawn from our simulation results: replicating the filter bubble phenomenon and exploring causal discovery in movie recommendation tasks. 
We acknowledge the difficulty of RQ2, and confine our discussion here to potential paths for future exploration.

\subsection{Filter Bubble Effect} 
\textbf{Motivation.} 
The filter bubble effect is a pervasive challenge in recommender systems \cite{VAE_filter_bubble, Explore_filter_bubble}. 
This issue emerges when algorithm-based recommenders predict which movies users might prefer based on user feedback loops \cite{Feedback_loop}, resulting in increasingly homogeneous recommended contents. 
Our primary goal is to assess Agent4Rec's ability to replicate the filter bubble phenomenon.

\smallskip\noindent\textbf{Setting.} 
To ensure fairness, we divide the MovieLens' movie pool into four equal parts, allowing the MF-based recommender to undergo four complete simulation rounds. 
In each round, recommended movies span up to 5 pages, with the MF-based recommender retrained after each simulation round.
We evaluate the filter bubble effect based on content diversity at the individual user level. Two metrics are employed: $\overline{P}_{\text{top1-genre}}$, which represents the average percentage of top-1 genres among recommended movies, and $\overline{N}_{\text{genres}}$, indicating the average number of genres recommended in each simulation.

\begin{figure}[t]
    \vspace{10pt}
    \includegraphics[width=0.8\columnwidth]{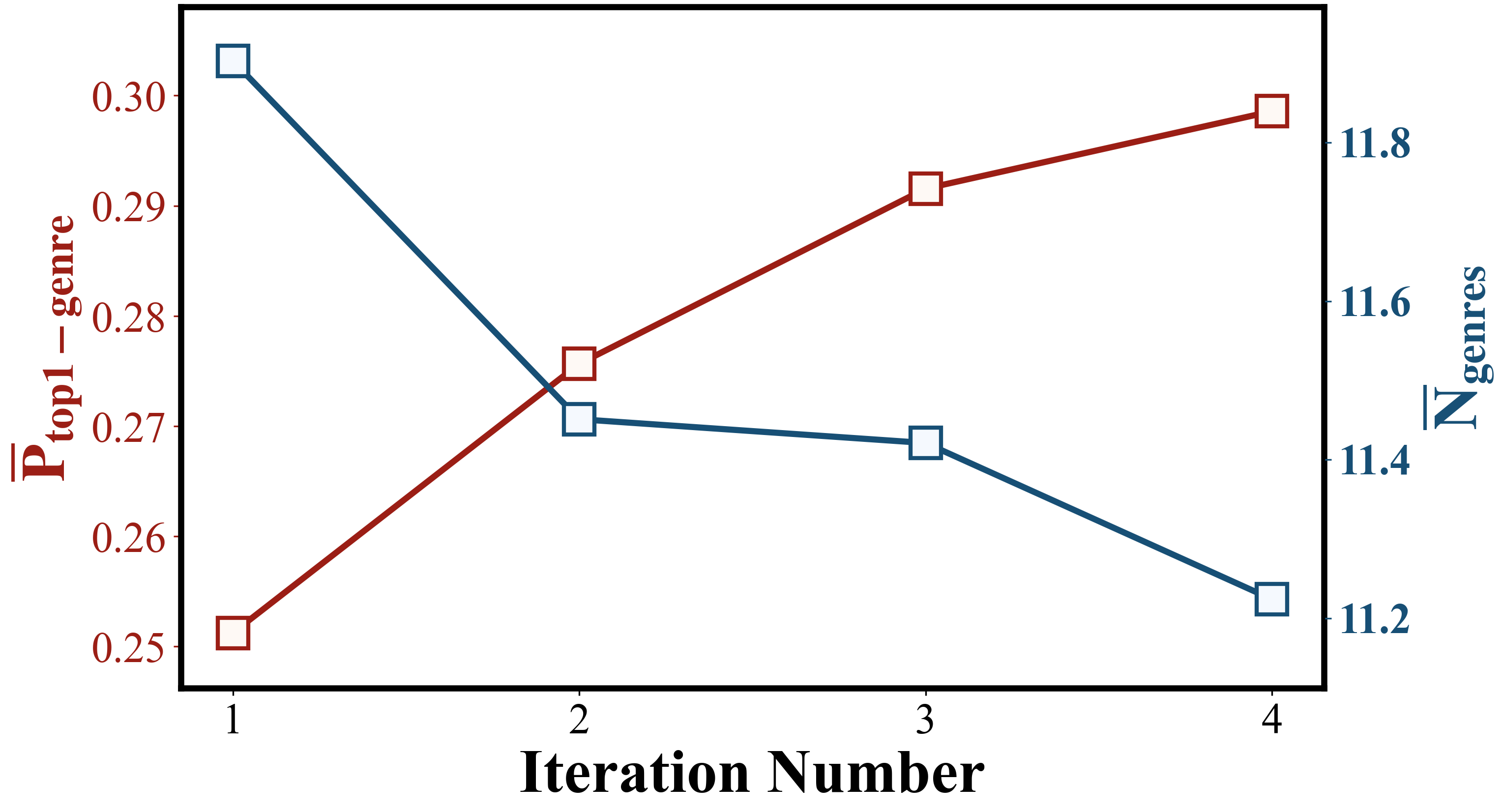}
    \vspace{-10pt}
    \caption{The simulation performance of Agent4Rec to emulate the filter bubble effect.}
    \vspace{-10pt}
    \label{fig:filter_bubble}
\end{figure}

\smallskip\noindent\textbf{Results.} 
Figure \ref{fig:filter_bubble} reveals that as the number of iterations increases, movie recommendations tend to be more centralized.
Specifically, the genre diversity, represented by $\overline{N}_{\text{genres}}$, decreases, while the dominance of the primary genre, denoted by $\overline{P}_{\text{top1-genre}}$, intensifies.
% See the details of the metrics in Appendix \ref{sec:filter_bubble}.
This result further validates Agent4Rec's capability to reflect the filter bubble effect, an issue commonly observed in real-world recommender systems.
% Figure \ref{fig:filter_bubble} indicates that the filter bubble phenomenon can be successfully replicated by our simulator. 
% As the recommender is continually influenced by agent feedback, it exhibits a propensity to recommend movies that align closely with users' preferences, as gleaned from their interaction history. 
% Consequently, this trend tends to result in a reduced diversity of movie genres within the recommender system. 
% Through this experiment, we present the potential to assess the effectiveness of strategies aimed at mitigating filter bubbles via a simulation platform like Agent4Rec, thus offering a cost-effective alternative to traditional online experiments.

\subsection{Discovering Causal Relationships}\label{sec:causal_discovery}

\begin{figure}[t]
    % \vspace{-10pt}
    \includegraphics[width=0.7\columnwidth, trim=190 90 230 0,clip]{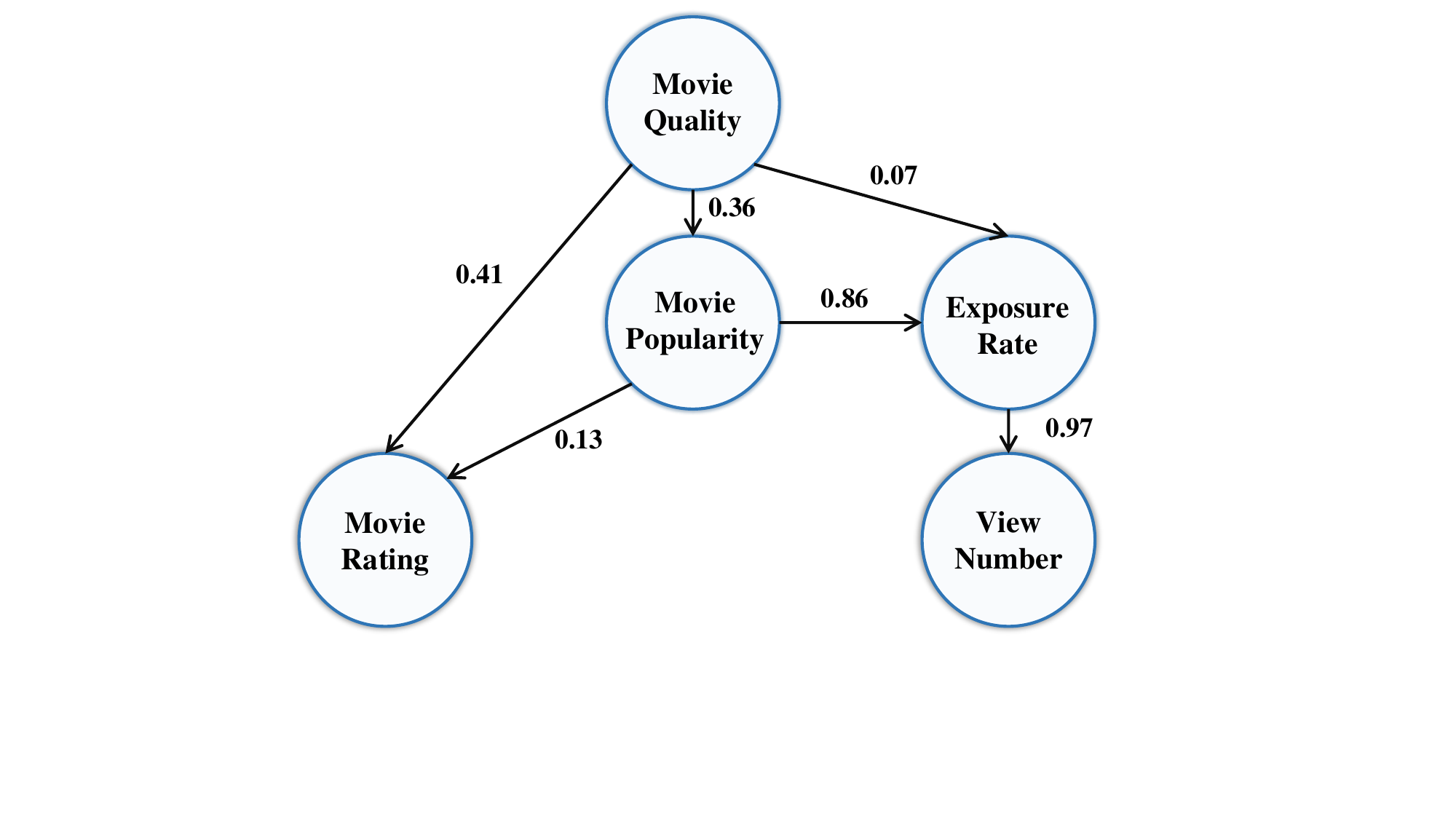}
    \vspace{-15pt}
    \caption{Learned causal graph among movie quality, popularity, exposure rate, view number, and movie rating.}
    \vspace{-10pt}
    \label{fig:causal_graph}
\end{figure}

\textbf{Motivation.} Causal discovery aims to infer a causal structure, often represented as a causal graph, from observational data. 
This technique is crucial to comprehending the underlying mechanisms of specific fields. 
A recommendation simulator can aid researchers with data collection and in addressing latent confounding issues. 
In light of this, a question arises: can Agent4Rec be instrumental in uncovering causal relationships in recommender systems?

\smallskip\noindent\textbf{Setting.} 
To understand the factors influencing movie ratings on MovieLens, for each movie, we collect data on four principal variables in addition to its rating simulated by agents: movie quality and popularity (sourced from the movie profile), exposure rate, and the number of times the movie is watched (sourced from the simulator). 
% See the detailed calculation for these factors in Appendix \ref{sec:causal_appendix}.
To probe the potential causal relationships within this simulated data, we employ the DirectLiNGAM algorithm \cite{LiNGAM}. 
This algorithm discovers a causal graph, \ie a weighted directed acyclic graph (DAG) in a linear system.

\smallskip\noindent\textbf{Results.} 
Based on the learned causal graph in Figure \ref{fig:causal_graph}, we can observe that: 
\begin{itemize}[leftmargin=*]
    \item \textbf{Movie quality and movie popularity are the \textit{causes} of movie ratings \cite{pearl2016causal}.} 
    Although movie quality contributes most to agent ratings, agent ratings are also slightly influenced by the popularity of movies. 
    This aligns with the inclination of humans to give high ratings to popular movies in real-world scenarios.
    \item \textbf{The feedback loop of amplifying the popularity bias is observed.} Highly popular movies receive increased exposure, resulting in more views by agents. When these popular items are introduced into the new training dataset, recommender systems are inclined to expose them more in the subsequent iterations, giving rise to what is known as popularity bias \cite{causal_survey, CausPref, InvCF}.
\end{itemize}

\section{Related Works}
In this section, we review two research lines of related works: LLM-empowered generative agents and recommender simulator.

\subsection{LLM-empowered Generative Agents}
AI agents are artificial entities that sense their environment, make decisions, and take actions \cite{Survey_Fudan}. Humans have long been pursuing intelligent agents to handle various tasks \cite{jennings1998roadmap, franklin1996agent, panait2005cooperative}.  Recently, with the remarkable capability of Large Language Model (LLM) demonstrated, a substantial body of research has emerged \cite{Survey_Renmin}. Generative Agent \cite{Generative_Agent} is a pioneer work that designs general agents equipped with memory, planning, and reflection abilities to simulate the human’s daily life. Building upon this universal framework, the following agent architecture can be bifurcated into task-oriented agents and simulation-oriented agents \cite{Survey_Fudan}.

The core objective of task-oriented agents is to execute predefined tasks established by humans \cite{Voyager, ChatDev, AutoGen, InteRecAgent, RecMind, RoCo}.  
Within this category, Voyager \cite{Voyager} integrates LLM into the Minecraft universe, endowing the in-game character with the capacity to effectively explore the virtual world. 
ChatDev \cite{ChatDev} introduces a novel agent collaboration paradigm in software development, harnessing group intelligence to streamline the development pipeline. AutoGen \cite{AutoGen} further extends this paradigm by defining more complex roles and facilitating multi-agent communication to tackle a diverse range of challenges.  In the realm of recommendation systems, RecMind \cite{RecMind} designs an LLM-powered autonomous recommender agent that possesses self-inspiring planning abilities and the capability to leverage external tools. InteRecAgent \cite{InteRecAgent} takes a step beyond by enhancing LLM-based agents with integrated components, yielding recommenders that are both explainable and conversational.

Simulation-oriented agents are geared towards replicating human behaviors in specific scenarios, thereby enabling the acquisition of valuable data and the exploration of social issues \cite{S3, Socially_alignment, RecAgent, WereWolf, Generative_Agent, AgentSims}. 
Among these agents, $S^{3}$ \cite{S3} leverages autonomous agents to simulate the dynamic evolution of public opinion on social platforms in response to trending social events. SANDBOX \cite{Socially_alignment} facilitates communication among a group of agents to address social issues thus providing ethically sound data for LLM fine-tuning. Differing from task-oriented agents in the field of recommendation, simulation-oriented agents seek to emulate user behaviors within recommender systems rather than focusing on the recommender. RecAgent \cite{RecAgent} attempts to integrate diverse user behaviors in recommendation environments, taking into account external social relationships. Focusing on simulating and evaluating user interactions with the recommenders, our framework Agent4Rec also falls into the category of simulation-oriented agents.

% In the realm of recommendation systems, there are few works available\cite{RecMind, InteRecAgent}. Among them, RecMind and InteRecAgent are task-oriented agents. RecMind \cite{RecMind} designs an LLM-powered autonomous recommender agent that possesses self-inspiring planning abilities and the capability to leverage external tools. InteRecAgent \cite{InteRecAgent} takes a step further by enhancing LLM-based agents with integrated components, yielding explainable and conversational recommender. Differing from the former works, simulation-oriented agents seek to emulate user behaviors within recommender systems rather than focusing on the recommender. RecAgent \cite{RecAgent} aims to simulate multi-type user behaviors in recommender system with external social relationships taken into consideration. Our framework Agent4Rec also falls into this category.

\subsection{Recommendation Simulator}
Recommendation simulator is a cornerstone in the field of recommendation systems \cite{reinforcement_survey_simulator, Feedback_loop, top_n_simulator, mindsim_simulator}. It diverges from conventional methods that rely on real-world user data, as it excels in replicating or simulating user interactions within recommender systems. This approach offers a cost-effective alternative to online environments and also presents a potential avenue for addressing prevalent challenges such as casual discovery and filter bubbles \cite{debias_simulator, adver_user_model_simulator}.
Early forays into recommendation simulators primarily serve as rich data sources for subsequent utilization, particularly in the realm of reinforcement learning \cite{Virtual-Taobao, RecSim, RecoGym}. For example, Virtual Taobao \cite{Virtual-Taobao} proposes a virtual environment based on real user interactions to simulate user behaviors on e-commerce platforms. RecSim \cite{RecSim} provides comprehensive toolkits for effectively simulating user behaviors in the setting of sequential recommendation. Also, RecoGym \cite{RecoGym} integrates both traditional recommendation algorithms and reinforcement learning framework, followed by evaluation with online and offline metrics. MINDSim \cite{mindsim_simulator} simulates user behaviors in a news website. But these traditional simulators all fall short in relatively simple rules, thus lacking flexibility and validity. Recently, LLM-empowered agents have demonstrated significant promise in approximating human-like intelligence, which showcases the considerable potential for recommendation simulators \cite{Survey_Fudan, Survey_Renmin}. RecAgent \cite{RecAgent} makes the primary attempt to construct a recommendation platform to integrate diverse user behaviors involving movie-watching, chatting, posting, and searching. Instead of behavior integration, we set our sights on in-depth exploration to emulate and evaluate user interactions with both rule-based and algorithm-based recommenders in recommender systems. 

\section{Limitations and Future Work}
Even though Agent4Rec offers a promising research direction in recommender system simulation, we recognize its potential limitations, risks, and challenges that require further exploration and in-depth investigation.

\begin{itemize}[leftmargin=*]
    \item \textbf{Datasource Constraints.}
    Agent4Rec is implemented exclusively utilizing offline datasets and is primarily constrained by two key factors. 
    First, LLMs necessitate prior knowledge regarding the recommended items, rendering most offline datasets --- with only IDs or lacking detailed item descriptions --- ill-suited for this task. 
    Furthermore, while online data undoubtedly aligns more naturally with simulators, providing an unbiased perspective to evaluate their effectiveness, acquiring such data poses a considerable challenge.
    \item \textbf{Limited Action Space.}
    The action space of Agent4Rec is currently limited, omitting critical factors that influence user decisions, such as social networks, advertising, and word-of-mouth marketing. 
    While this simplification facilitates a reliable evaluation of LLM-empowered agents under simple scenarios, it also introduces a gap in real-world user decision-making processes. 
    A key direction for our future work is to encompass a wider spectrum of influential factors to better capture the multifaceted nature of user behaviors, ensuring the simulations are more universally representative of recommendation scenarios.
    \item \textbf{Hallucination in LLM.}
    Occasional hallucinations have been observed in simulations, such as the LLM failing to accurately simulate human users providing unfavorable ratings to adopted items, fabricating non-existent items and rating them, or not adhering to the required output format. 
    Such inconsistencies can lead to inaccurate simulation outcomes. 
    In light of these observations, our future goal is to fine-tune an LLM specifically for simulating user behavior in recommendation scenarios to enhance the simulator's stability and precision.
\end{itemize}

% Dataset Limitation - MovieLens, online dataset
% Action space limitation - No social networks, mouth marketing, chatting
% LLM's inherent bias - Hallucinations, randomness

%实验中发现的limitation

%1.Fail to simulate complicated emotion evolution. 在电影评分中不给1-2分，即不具备在判断电影符合Taste并选择观看后打出低分的能力。在现实生活中打出低分的理由有很多，如不满意电影的某些方面或感到不符预期。本质上是在细粒度的情感模拟方面做的不太好（good at factual but not so good at emotion, but the decisions in recommender system highly rely on user's emotion）。

%2.LLM has Hallucinations. During reaction to recommended movie list, sometimes respond to movies that don't appear in the recommender system.

%3.LLM sometimes fails to execute user's requirements. Sometimes disobey our requirements to output the response with a certain format, causing the failure of regular expression, leading to incomplete behavior records.

%4.相同setting下， 两次实验结果可能不一样，不稳定。但这也符合真实用户做选择时的“一时兴起”

%5.LLM needs to be tailored for recommendations to eliminate some prior knowledge. 比如，除非用prompt高度限制或引入疲劳度进行限制，否则用户即使连续的感到unsatisfied，也会选择继续看下一页，这与真实用户的表现有gap。

%6.受限于movielens数据集

% \newpage

\begin{acks}
This research is supported by the National Science and Technology Major Project (2023ZD0121102), National Natural Science Foundation of China (92270114), the National Research Foundation, Singapore under its Industry Alignment Fund – Pre-positioning (IAF-PP) Funding Initiative, and the NExT Research Center. 
Any opinions, findings, conclusions, or recommendations expressed in this material are those of the author(s) and do not reflect the views of National Research Foundation, Singapore.
\end{acks}


\bibliographystyle{ACM-Reference-Format}
% \newpage
\bibliography{2024_www}

% \newpage
% \input{chapters/7_appendix}

\end{document}